\documentclass[modern]{aastex61}

\usepackage{rotating}
\usepackage{mathtools}

\newcommand{\lalpha}{Lyman-$\alpha$}
\newcommand{\angstrom}{\textup{\AA}}

\submitjournal{ApJL}

\shorttitle{A new mechanism for atmospheric escape}
\shortauthors{Garc\'ia Mu\~noz \& Schneider}

\begin{document}

\title{Rapid escape of ultra-hot exoplanet atmospheres driven by hydrogen Balmer absorption}

\correspondingauthor{Antonio Garc\'ia Mu\~noz}
\email{garciamunoz@astro.physik.tu-berlin.de, tonhingm@gmail.com}

\author{A. Garc\'ia Mu\~noz}
\affiliation{Zentrum f\"ur Astronomie und Astrophysik, Technische Universit\"at Berlin, \\
Hardenbergstrasse 36, D-10623, Berlin, Germany}

\author{P.C. Schneider}
\affiliation{Hamburger Sternwarte, Gojenbergsweg 112, 21029, Hamburg, Germany}

\begin{abstract}
Atmospheric escape is key to explaining the long-term evolution of planets in our Solar System 
and beyond, 
and in the interpretation of atmospheric measurements. 
Hydrodynamic escape is generally thought to be driven by
the flux of extreme ultraviolet photons 
that the planet receives from its host star. 
Here, we show that the escape from planets orbiting hot stars proceeds through a 
different yet complementary process: drawing its energy from the intense near ultraviolet emission 
of the star that 
is deposited 
 within an optically thin, high-altitude atmospheric layer 
 of hydrogen excited into the lower state of the Balmer series. 
The ultra-hot exoplanet KELT-9b likely represents the first known instance of this 
Balmer-driven escape. 
In this regime of hydrodynamic escape, the near ultraviolet emission from the star 
is more important at determining the planet mass loss
than the extreme ultraviolet emission, 
and uncertainties in the latter become less critical.
Further, we predict that gas exoplanets around hot stars may experience 
catastrophic mass loss when they are less massive than 1--2 Jupiter masses 
and closer-in than KELT-9b, 
thereby challenging the paradigm that all large
exoplanets are stable to atmospheric escape. 
We argue that
extreme escape will affect the demographics of close-in exoplanets orbiting hot stars. 
\end{abstract}

\keywords{ultra-hot Jupiter --- thermosphere --- NLTE --- hydrodynamics --- atmospheric escape}

\section{Introduction} \label{sec:intro}

Planet atmospheres subject to strong stellar irradiation undergo hydrodynamic escape that  
may affect the planets' bulk properties if sustained over Gigayears, 
especially for the smaller, lower-mass planets  
\citep{lammeretal2008, tian2015, 
zahnlecatling2017, owen2019}. 
This idea is supported by the statistics of the $>$4,000 exoplanets 
discovered to date, which show that specific combinations 
of planet size and stellar irradiation are underrepresented 
-- a finding consistent with significant planet mass evolution
\citep{lopezetal2012,fultonetal2017,jinmordasini2018}.
On the other hand,
hydrodynamic escape is thought to affect minimally the evolution of Jupiter-mass and heavier
planets. 
Also, in some instances atoms of suspected atmospheric origin 
have been detected at altitudes near or beyond the planets' Roche lobe
\citep{vidalmadjaretal2003,vidalmadjaretal2004,benjaffel2007,fossatietal2010, 
lecavelierdesetangsetal2010, linskyetal2010, jensenetal2012, 
kulowetal2014, ballesterbenjaffel2015, 
ehrenreichetal2015, bourrieretal2018,  
salzetal2018, spakeetal2018, singetal2019}, indicating that they 
are no longer gravitationally bound. 
\\

The generally accepted view of hydrodynamic escape in hydrogen-dominated atmospheres 
is that it is driven by stellar extreme ultraviolet (EUV; wavelengths $\lambda$$<$912 {\AA}) photons 
deposited in the planet's thermosphere 
\citep{lammeretal2003, yelle2004, tianetal2005, 
garciamunoz2007, murrayclayetal2009, koskinenetal2013,ionovetal2014, guobenjaffel2016, 
salzetal2016}. 
In that view, ground state hydrogen H(1) (principal quantum number $n$ in parentheses)
plays the fundamental role of absorbing the incident photons in the 
Lyman continuum ($n$=1$\rightarrow$$\infty$) that ultimately heat the atmosphere. 
It is also tacitly assumed that no other gas competes with H(1) 
in terms of stellar energy absorption at these or other wavelengths in the thermosphere, 
even when absorption by metals at wavelengths longer than the Lyman continuum threshold
is considered \citep{garciamunoz2007,koskinenetal2013}. 
Our work challenges these ideas for, at least, the case of exoplanets orbiting hot stars.
\\

The discovery of exoplanets around hot stars of effective temperatures 
$T_{\rm{eff}}$$>$7,500 K 
\citep{colliercameronetal2010,gaudietal2017, lundetal2017, 
talensetal2018} 
and the detection of absorption by atomic gases in their atmospheres 
\citep{casasayasbarrisetal2018,hoeijmakersetal2018,yanhenning2018,cauleyetal2019} 
has opened up new opportunities to test our understanding of hydrodynamic 
escape and in turn of exoplanet evolution. 
The hottest of these stars ($T_{\rm{eff}}$$>$8,200 K) 
present high levels of near-ultraviolet (NUV; for convenience  
loosely defined here as having $\lambda$$<$4,000 {\AA}, 
although in strict terms that also includes the middle and far ultraviolet)
emission but are not expected to be particularly strong EUV emitters \citep{fossatietal2018}. 
The planets transiting two of such hot stars, KELT-9b and MASCARA-2b/KELT-20b, 
exhibit strong absorption 
in the H-$\alpha$ (6565 {\AA}; $n$=2$\rightarrow$$n$=3) and 
-$\beta$ (4863 {\AA}; $n$=2$\rightarrow$$n$=4) lines of the hydrogen 
Balmer series 
\citep{yanhenning2018,casasayasbarrisetal2018,cauleyetal2019}.
It is thus conceivable that Balmer continuum absorption
($n$=2$\rightarrow$$\infty$; $\lambda$$<$3646 {\AA}), 
facilitated by large column abundances of thermospheric H(2), will tap into the 
enormous reservoir of energy in the stellar NUV emission. 
This represents a previously unconsidered source of energy 
to drive hydrodynamic escape that potentially outpaces the EUV-driven escape.
\\

The condition for this to occur is 
$\tau_{\rm{BaC}}$$F^{\star}_{\rm{BaC}}$/$F^{\star}_{\rm{LyC}}$$\gtrsim$1, where 
$\tau_{\rm{BaC}}$ is the optical thickness at a representative Balmer continuum wavelength
from the atmospheric top to the lower thermosphere, 
and $F^{\star}_{\rm{BaC}}$/$F^{\star}_{\rm{LyC}}$ is the ratio of
wavelength-integrated stellar emission energies
in the Balmer and Lyman continua. 
The condition derives from estimating 
the energy deposited at each wavelength range
$\propto$$F^{\star}$$\left[1-\exp(-\tau)  \right]$, and realizing that
the thermosphere is optically thick at EUV wavelengths but 
thin in the NUV, i.e. $\tau_{\rm{LyC}}$$\gg$1
but $\tau_{\rm{BaC}}$$\ll$1.
We estimate for the star KELT-9 (at 1 Astronomical Unit; see below), 
$F^{\star}_{\rm{LyC}}$=3.8 and $F^{\star}_{\rm{BaC}}$=2.9$\times$10$^7$ erg s$^{-1}$ cm$^{-2}$, 
which results in the ratio $\sim$7.5$\times$10$^6$. 
An optical thickness $\tau_{\rm{BaC}}$$\gtrsim$1.3$\times$10$^{-7}$ will turn 
the stellar NUV emission into the main source of energy deposited in KELT-9b's thermosphere.
For reference, typical values for the Sun
($T$$_{\rm{eff}}$=5,800 K) are 4.4 and 7.1$\times$10$^4$ erg s$^{-1}$ cm$^{-2}$, respectively,
and $F^{\star}_{\rm{BaC}}$/$F^{\star}_{\rm{LyC}}$$\sim$1.6$\times$10$^4$, 
as estimated from the ATLAS1/SOLSPEC solar irradiance spectrum of \citet{thuillieretal2004} 
presented by \citet{schoelletal2016} and available at 
http://projects.pmodwrc.ch/solid/.
\\

\section{Model} \label{sec:model}

We built a model of KELT-9b's hydrogen atmosphere
to investigate whether the detected H-$\alpha$ and -$\beta$ lines
are the smoking guns for significant stellar energy absorption in the hydrogen 
Balmer continuum that could result in a vigorous hydrodynamic escape. 
The model solves the conservation equations of the 
expanding thermosphere \citep{garciamunoz2007}, and incorporates 
a formulation for Non-Local Thermodynamic Equilibrium (NLTE) 
in an atomic hydrogen gas \citep{munafoetal2017}, 
thus coupling the population of hydrogen states with the radiation field and the 
hydrodynamics. 
The importance of NLTE is well documented in stellar astrophysics but remains,
save for a few exceptions 
\citep{christieetal2013,menageretal2013,huangetal2017}, 
poorly explored for exoplanets. 
\\

The conservation equations are complemented with boundary conditions. 
The bottom boundary is placed at a pressure  $p$$\sim$0.1 dyn cm$^{-2}$ 
(corresponding to one planet radius $R_{\rm{p}}$/$R_{\rm{J}}$=1.89; $R_{\rm{J}}$ is
Jupiter's radius), where we impose the temperature. 
We first tried with a temperature of 4,600 K, 
which is consistent with the occultation brightness 
measured for KELT-9b \citep{gaudietal2017} and that probably arises from the 
planet's lower atmosphere. 
Our full-model calculations revealed however that the temperature at pressures $p$$\le$0.1 dyn cm$^{-2}$ is 
 dictated by radiation from higher altitudes and that the choice of bottom temperature in
the model is of minor importance. 
The simulations presented here were carried out with a bottom
boundary temperature of 8,000 K that minimizes the overshoot in the first few points
of the spatial grid. This choice has no bearing on 
the overall solution for the conditions that we explored (see Appendix). 
The velocity and volume mixing ratios at the  bottom boundary 
are extrapolated from inside the model domain. 
This type of floating conditions for the abundances prevents strong gradients in composition. 
The top boundary is placed at a radial distance $r$/$R_p$=2.5, near the Lagrangian L1 
point in the substellar direction ($r_{\rm{L1}}$/$R_p$=2.65), where 
 we impose that the flow is supersonic.
The atmosphere is irradiated from the top.  
The system of equations and boundary conditions is solved numerically following the methods in 
\cite{garciamunoz2007} until the steady state is reached for all the variables in the 
hydrodynamic and NLTE problems. 
We calculate the mass loss rate from $\dot{m}$=$\pi$$\rho u r^2$, 
where $\rho$ and $u$ are the mass density and mass average velocity of the gas as a whole, 
respectively. 
\\

At the high temperatures of KELT-9b's thermosphere all molecules are 
dissociated \citep{kitzmannetal2018,lothringeretal2018}. 
Thus, our adopted chemical scheme is based on a hydrogen atom model, and 
includes neutral atoms, protons and electrons. 
For the neutral atom, 7 bound states are considered: the ground state plus 6 excited states. 
This makes a total of 9 pseudo-species, each of them considered 
separately in the chemical scheme.
The 4 lower bound states are resolved by their 
principal ($n$=1$-$2), orbital ($\ell$) and total angular momentum ($J$) quantum numbers. 
They are: $1s$ $^2S_{1/2}$, $2p$ $^2P^*_{1/2}$, $2s$ $^2S_{1/2}$, $2p$ $^2P^*_{3/2}$. 
The 3 upper bound states are identified solely by their principal quantum numbers  
($n$=3$-$5) after averaging over fine-structure details.  
The hydrogen atom model is complex enough to treat the most interesting phenomena in
KELT-9b's thermosphere, which involve preferentially the lower energy states. 
The 9 pseudo-species in our hydrogen atom model interact through 97 channels, namely: 
electron-collision excitation (21 channels), de-excitation (21), ionization (7), 
and three-body recombination (7);
proton-collision mixing (4); 
photoionization (7); radiative recombination (7);  
photoexcitation (2); spontaneous emission (21). 
\\

To determine the radiation field we solve the radiative transfer equation:
\begin{equation*}
\frac{dI_{\lambda}}{ds}=-\kappa_{\lambda}I_{\lambda} + \varepsilon_{\lambda}, 
\label{rte_eq}
\end{equation*}
which includes terms for absorption and emission from the 
bound-bound (BB), bound-free/free-bound (BF/FB) and free-free (FF)
radiative transitions in the NLTE scheme. Thus, 
$\kappa_{\lambda}$=$\kappa^{\rm{BB}}_{\lambda}$+$\kappa^{\rm{BF}}_{\lambda}$+$\kappa^{\rm{FF}}_{\lambda}$  
and 
$\varepsilon_{\lambda}$=$\varepsilon^{\rm{BB}}_{\lambda}$+$\varepsilon^{\rm{FB}}_{\lambda}$+$\varepsilon^{\rm{FF}}_{\lambda}$ 
for the absorption and emission coefficients, respectively. 
$I_{\lambda}$$(\mathbf{x,s})$ is the radiance, and $d$/$ds$ its spatial derivative. 
For simplicity, the only BB transition considered in the radiative transfer problem  
is {\lalpha}, which means that in the solution to the above equation we adopt 
$\kappa^{\rm{BB}}_{\lambda}$$\approx$$\kappa^{\rm{Ly}-\alpha}_{\lambda}$ and
$\varepsilon^{\rm{BB}}_{\lambda}$$\approx$$\varepsilon^{\rm{Ly}-\alpha}_{\lambda}$ and
their wavelength-dependent descriptions.
However, for the net energy emission rate $\Gamma$  
we consider all of the BB transitions 
under the assumption that all of them except {\lalpha} do emit 
but do not absorb. A more comprehensive treatment of BB transitions will 
foreseeably enhance the proposed mechanism (see Appendix). 
We solve the radiative transfer equation over a spectral grid of 
varying resolution and a total of 751 spectral bins. 
Particular emphasis is placed on resolving the {\lalpha} line, and indeed
the bin size near the line core is as small as a fraction of a thermal broadening width.
The radiative transfer equation is solved 
in a plane-parallel atmosphere that mimics our model atmosphere along the substellar line. 
Neglecting curvature effects is an acceptable approximation, especially near
the model bottom boundary, which is where radiative effects are more important.
\\

The radiative transfer equation dictates how much radiation is deposited and where.
We define the net energy emission rate from radiative processes
$\Gamma$ 
by integrating the radiative transfer equation over wavelength and solid angle:
\begin{equation*}
\Gamma(\mathbf{x})=\int \int 
[-\kappa_{\lambda}(\mathbf{x})I_{\lambda}(\mathbf{x,s}) + \varepsilon_{\lambda}(\mathbf{x})]
d\Omega(\mathbf{s}) d\lambda. 
\label{lambda_eq}
\end{equation*}
$\Gamma$$>$0 and $<$0 represent net local cooling and heating, respectively.
We solve both the problems of direct stellar (non-diffuse) radiation and 
of diffuse radiation. Their solutions provide 
$I_{\lambda}(\mathbf{x,s})$=$I_{\lambda}^{\star}(\mathbf{x,s})$+$I_{\lambda}^{d}(\mathbf{x,s})$. 
$\Gamma$ appears in the energy conservation equation as a non-local energy 
source (or sink) that connects the gas over a range of altitudes. 
\\

The specifics of the stellar spectrum play a fundamental role in KELT-9b's thermospheric
structure.
For our work, we used the PHOENIX LTE spectrum \citep{husseretal2013} for $T_{\rm{eff}}$=10,200 K, 
resulting in the aforementioned $F^{\star}_{\rm{LyC}}$ and $F^{\star}_{\rm{BaC}}$. 
The earlier studies of KELT-9b proposed that the planet is 
subject to strong EUV irradiation from its host star \citep{gaudietal2017}.  
This idea has been revised \citep{fossatietal2018}, noting that
intermediate-mass stars hotter than $\sim$8,250 K  
lack a chromosphere and corona and consequently their EUV emissions are small or moderate. 
The latter work estimates that the EUV irradiation received by
KELT-9b on its orbit is probably on the order of $\sim$4,000 erg s$^{-1}$cm$^{-2}$, consistent
with our adopted value. 
Most of the stellar EUV emission occurs near 
the Lyman-continuum edge, which entails that it is deposited over a narrow altitude range 
in KELT-9b's thermosphere.

\section{Results} \label{sec:results}

We define our fiducial model for KELT-9b as having 
a mass $M_{\rm{p}}$/$M_{\rm{J}}$=2.88 \citep{gaudietal2017}. 
In this model (black curves in Fig. \ref{mypanel_fig}), 
the stellar EUV energy is deposited near the 
$\tau_{\rm{LyC}}$=1 level ($p$$\sim$3$\times$10$^{-3}$ dyn cm$^{-2}$; $r$/$R_{\rm{p}}$$\sim$1.27) 
and triggers the ionization of the gas.
The transition between H(1) and H$^+$ as the main form of hydrogen occurs 
at $p$$\sim$1.7$\times$10$^{-2}$ dyn cm$^{-2}$ ($r$/$R_{\rm{p}}$$\sim$1.10), near the bottom boundary. 
Part of the diffuse Lyman continuum radiation arising from the recombining plasma 
reaches the lower thermosphere (i.e. deeper than the $\tau_{\rm{LyC}}$=1 level), 
further ionizing and heating the deeper atmospheric layers. 
Since the local plasma is optically thick at Lyman continuum wavelengths, 
the process of ionization and recombination occurs multiple times. 
As radiation diffuses and temperatures increase in the lower thermosphere, 
the population of H(2) also increases -- and eventually becomes large enough as to 
intercept a significant amount of stellar NUV photons -- 
which further increases the local temperature. 
This multi-step process explains the high temperatures of $\sim$10,000 K 
reached throughout the lower thermosphere and that override our bottom boundary condition
for temperature. 
The temperature profile (and the mass loss rate; see below) in our full-model calculations
differs significantly from the predictions when 
we impose $F^{\star}$($\lambda$$>$912 {\AA})$\equiv$0 in the model 
(dashed-dotted curve, Fig. \ref{mypanel_fig}). For the latter conditions, 
diffuse radiation produces a bulge of temperature in the lower thermosphere. 
The bulge vanishes when we additionally turn off the NLTE scheme (not shown), in which case
the lower thermosphere develops a temperature 
minimum similar to those predicted in published works of planets orbiting cooler stars
\citep{garciamunoz2007}. 
For KELT-9b, 
the absorption of stellar NUV energy by excited hydrogen throughout its thermosphere
washes out such a minimum.
\\

\begin{figure*}
   \centering
   \includegraphics[width=9.cm]{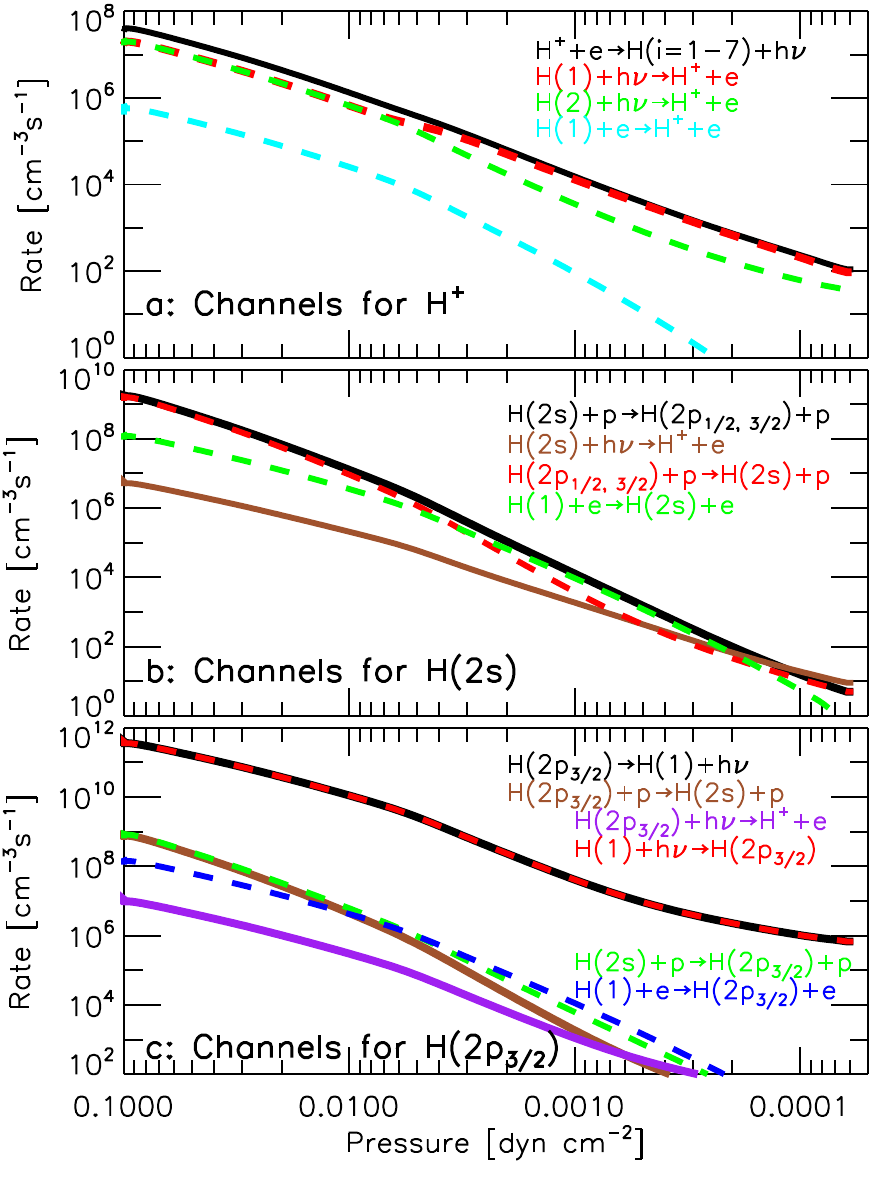}
      \caption{\label{XX_fig} Channels of formation (dashed curves) and destruction (solid curves) for
      H$^{+}$ (\textbf{a}), H(2$s$) (\textbf{b}) and H(2$p_{3/2}$) (\textbf{c}). 
      In \textbf{a}, the radiative recombination channel contains products in all possible 
      bound states, and H(2) refers to all the states with principal quantum number $n$=2.
      In \textbf{b}-\textbf{c}, p actually refers to collisions with both protons and electrons.
      }
   \end{figure*}

Our model shows that the H(1) and H(2) abundances are closely tied. 
In particular, 
the densities of the $2p$ states are largely dictated by
photoexcitation from the ground state and subsequent radiative decay.
In the lower thermosphere, where the densities are higher, 
the  $2s$ $^2S_{1/2}$ state reaches an equilibrium through proton collisions with
the $2p$ $^2P^*_{1/2}$ and $2p$ $^2P^*_{3/2}$ states. 
The number density ratio H(2):H(1) in most of the thermosphere is
a few times 10$^{-8}$, 
notably smaller than the LTE prediction ($\sim$3$\times$10$^{-5}$ at 10,000 K). 
Unsurprisingly, LTE fails to provide a realistic description of the rarefied atmosphere. 
The formation of H$^{+}$ proceeds through 
photoionization of H(1) and, interestingly, also of H(2). 
Indeed, although the H(2) abundances are small, the corresponding 
photoionization coefficients are orders of magnitude larger than for H(1). 
The neutralization of H$^+$ occurs mainly through radiative recombination, 
which populates the excited states that eventually cascade into H(1). 
Figure (\ref{XX_fig}) shows the reaction rates for the main formation and destruction 
channels of H$^+$ and the bound states $2s$ $^2S_{1/2}$ and $2p$ $^2P^*_{3/2}$. 
\\

\begin{figure*}[h]
   \centering
   \includegraphics[width=18.cm, angle=0]{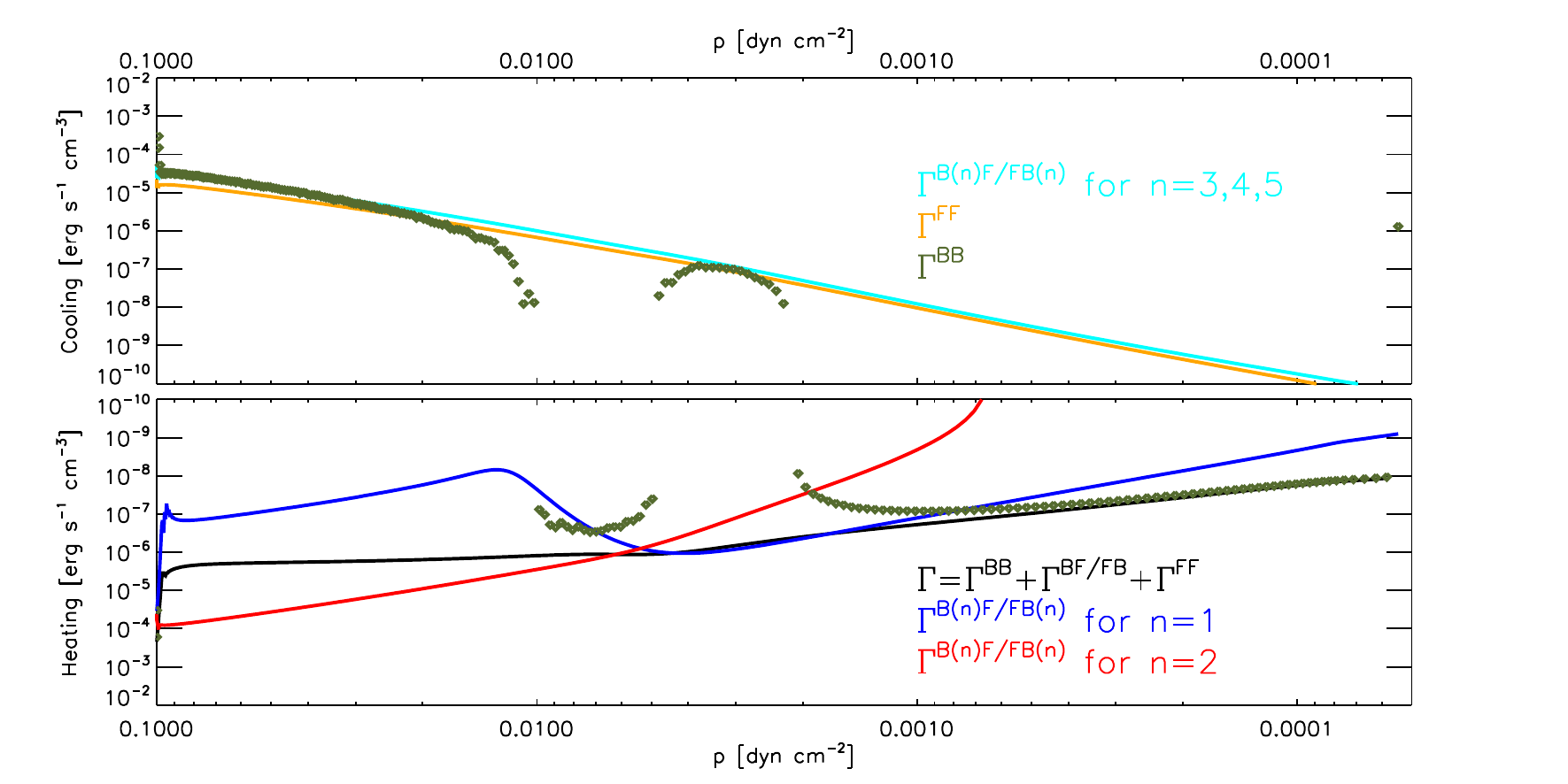}
      \caption{\label{gamma_fig} 
      For the fiducial model, net energy emission rates. 
      As defined in the text, $\Gamma$
      is calculated at each altitude by integration over wavelength and solid angle 
      with the emission and absorption coefficients specific to the designated
      BB, BF/FB or FF transition.
      Each component of $\Gamma$ can be either positive (cooling; in which case it 
      appears in the upper panel) or negative (heating; it appears in the lower panel), and they
      can transition from positive to negative (or vice versa) as the specific 
      mechanism switches from cooling to heating (or vice versa).
      For BF/FB transitions, there are three separate curves for processes involving
      the principal quantum numbers $n$=1, 2, and 3--5. The net heating 
      ($\Gamma^{B(2)F/FB(2)}$$<$0; red curve) that occurs through the photoionization
      and subsequent radiative recombination 
      H(2)+$hc$/$\lambda$ ($\lambda$$<$3,646 {\AA})$\leftrightarrow$ H$^+$+e dominates 
      the energy budget in the lower thermosphere. 
      }
   \end{figure*}

The H(2):H(1) ratio  and the large scale heights  
ensure that the stellar energy deposited in the thermosphere at Balmer continuum 
wavelengths eventually becomes much larger than in the Lyman continuum.
This is confirmed in Fig. (\ref{gamma_fig}), which presents a breakdown of the contributions
from BB, BF/FB and FF transitions to the net energy emission rate, 
$\Gamma(\mathbf{x})=\Gamma^{\rm{BB}}(\mathbf{x})+\Gamma^{\rm{BF/FB}}(\mathbf{x})+\Gamma^{\rm{FF}}(\mathbf{x})$. 
Heating in the lower thermosphere is dominated by the net effect of photoionization and
radiative recombination from and into the  bound state H(2).

\section{Discussion and summary} 

\begin{figure}[h]
   \centering
   \includegraphics[width=10.cm]{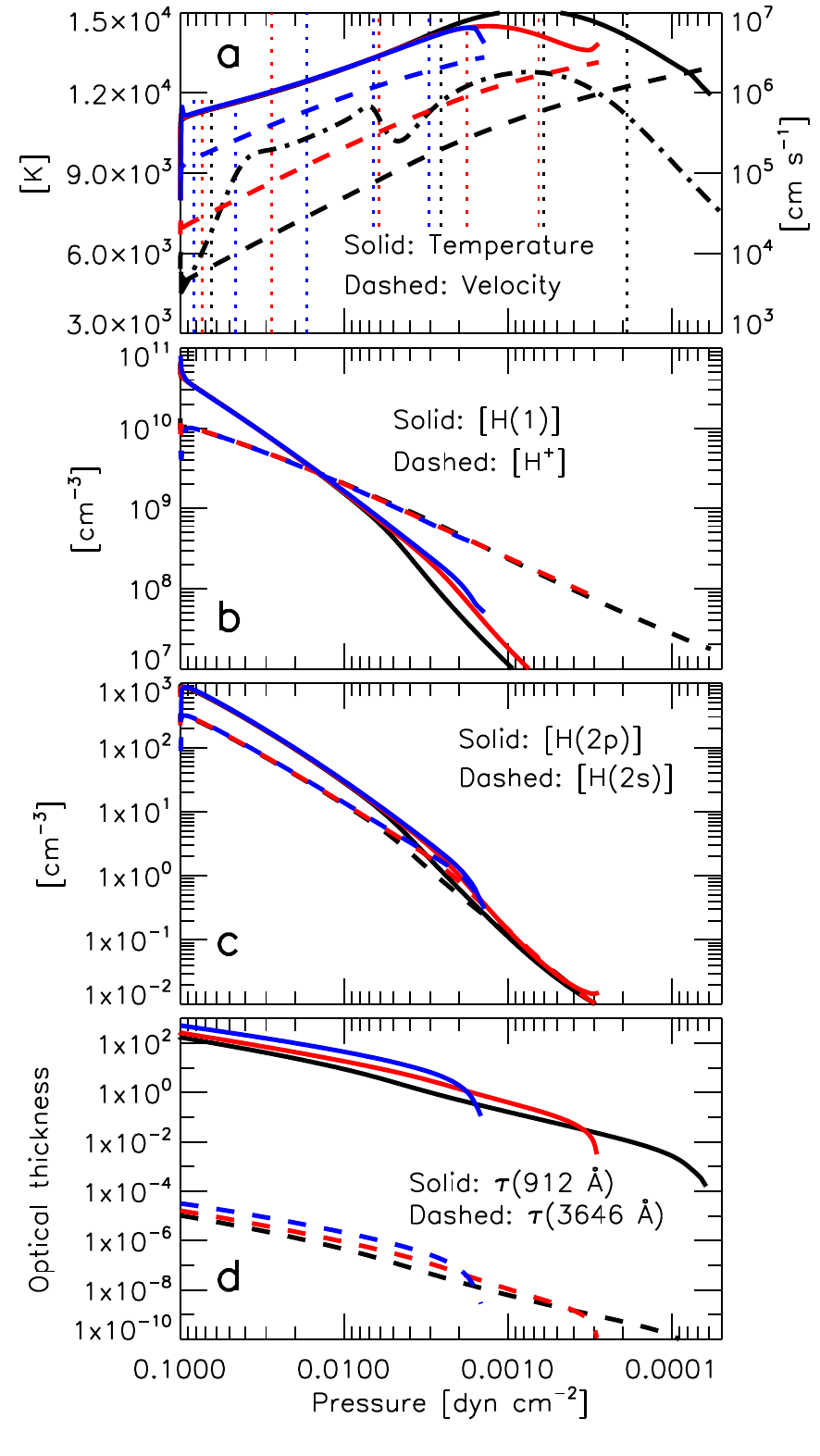}
   \caption{\label{mypanel_fig}
   Model profiles. Adopted planet masses:  
   $M_{\rm{p}}$/$M_{\rm{J}}$=2.88 (black), 2.04 (red), and 1.20 (blue).
   \textbf{(a)} Temperature and velocity; 
   \textbf{(b)} H(1) and H$^+$ number density; 
   \textbf{(c)} H(2p) (=H(2p$_{1/2}$)+H(2p$_{3/2}$)) and H(2s) number density; 
   \textbf{(d)} Optical thickness $\tau_{r \rightarrow \infty}$ at the edge of the Lyman 
   ($<$912 {\AA}) and Balmer ($<$3646 {\AA}) continua.
   In \textbf{(a)} the dotted vertical lines refer to $r$/$R_{\rm{p}}$=1.02, 1.1, 1.3, 1.6, 2.
   Each model is run from $r$/$R_{\rm{p}}$=1 to 2.5.
   In \textbf{(a)}, the dashed-dotted curve shows the temperature profile for a simulation 
   with $M_{\rm{p}}$/$M_{\rm{J}}$=2.88 and $F^{\star}$($\lambda$$>$912 {\AA})$\equiv$0.
   }
   \end{figure}

The mass determination of exoplanets orbiting hot stars is inherently difficult  
and indeed KELT-9b's mass is rather uncertain
($M_{\rm{p}}$/$M_{\rm{J}}$=2.88$\pm$0.84; 1-$\sigma$)  \citep{gaudietal2017}. 
We take this uncertainty as a motivation to explore the plausible conditions 
in the atmospheres of similar planets having a range of masses.
Adopting $M_{\rm{p}}$/$M_{\rm{J}}$=2.04 (red curves, Fig. \ref{mypanel_fig}) and 1.20 (blue curves), 
it is seen that the atmosphere becomes progressively extended. 
This is especially important in the Balmer continuum, as 
the increasing H(2) column provides a means
of depositing more of the stellar NUV energy at high altitudes. 
Indeed, $\tau_{\rm{BaC}}$$\sim$10$^{-7}$ is reached at 
$p$$\sim$4.3$\times$10$^{-3}$ dyn cm$^{-2}$ ($r/R_{\rm{p}}$$\sim$1.23) 
in our fiducial model and at $p$$\sim$2.1$\times$10$^{-3}$ dyn cm$^{-2}$
($r/R_{\rm{p}}$$\sim$2.20) for $M_{\rm{p}}$/$M_{\rm{J}}$=1.20.
\\

\begin{figure}[h]
   \centering
   \includegraphics[width=10.cm]{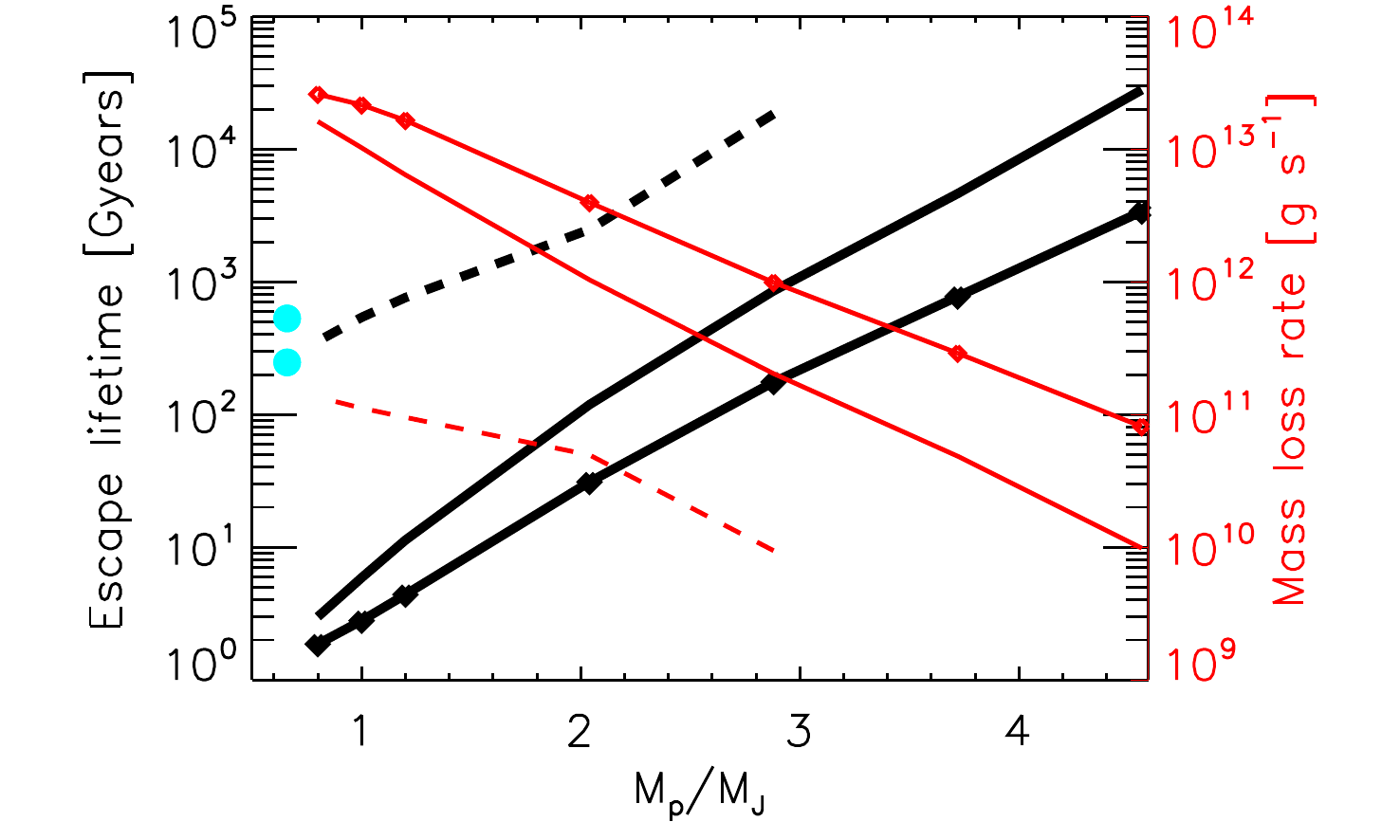}
   \caption{\label{mylifetime_fig} 
   Escape lifetimes $t_{\rm{esc}}$ and mass loss rates $\dot{m}$. 
   Each set of curves corresponds to a model configuration specific to KELT-9b. 
   Dashed: $F^{\star}$($\lambda$$>$912 {\AA})$\equiv$0. 
   Solid: Full stellar spectrum, orbital distance $a$=0.035 AU.
   Diamonds: Full stellar spectrum, $a$=0.025 AU.
   Cyan circles refer to simulations for HD209458b, 
   with the low mass loss rate referring to 
   $F^{\star}$($\lambda$$>$912 {\AA})$\equiv$0, and the
   high mass loss rate referring to the full stellar spectrum.   
   }
   \end{figure}

The predicted mass loss rates $\dot{m}$ are very sensitive to the adopted $M_{\rm{p}}$ 
(Fig. \ref{mylifetime_fig}; black solid curves). 
Our full-model predictions for $\dot{m}$ are larger by up to two orders of magnitude than what is predicted if 
only the  stellar EUV spectrum is considered 
(i.e. if we set $F^{\star}$($\lambda$$>$912 {\AA})$\equiv$0; black dashed curves).
Defining a escape lifetime as $t_{\rm{esc}}$=$M_{\rm{p}}$/$\dot{m}$, 
it is seen that $t_{\rm{esc}}$ becomes on the order of Gigayears for 
$M_{\rm{p}}$/$M_{\rm{J}}$$\lesssim$1.20. The real time for the planet to fully lose its 
atmosphere will probably be shorter because as the planet 
loses mass its gravitational potential becomes shallower, 
provided that the planet mass evolves much more rapidly than its size.
In such conditions, the planet will foreseeably lose its atmosphere in less than a 
Gigayear. 
This interesting possibility should be tested with a more comprehensive calculation
that considers the interior structure of the planet as well as the coupling between the
lower and upper atmospheres.
Further, our model shows that $\dot{m}$ ($t_{\rm{esc}}$) will be much higher (lower) for closer-in orbits 
(Fig. \ref{mylifetime_fig}; black diamonds are for an orbital distance of
0.025 AU$<$0.035 AU for current KELT-9b). 
These considerations set important constraints on the stability of exoplanets 
orbiting hot stars that can be tested by ongoing surveys of transiting
exoplanets. 
In particular, Balmer-driven escape may help explain the well-known 
lack of large exoplanets on short-period orbits  -- the so-called evaporation desert
\citep{owen2019,lopezetal2012,fultonetal2017,jinmordasini2018} 
-- at least around hot stars. In this Balmer-driven regime of hydrodynamic escape, 
the stellar EUV radiation becomes of second order importance, 
and uncertainties in the EUV flux are not critical for the prediction of $\dot{m}$. 
The foregoing discussion highlights the importance of physically-motivated 
hydrodynamic escape models 
as opposed to parameterizations based on e.g. energy-limited escape
to investigate the mass loss from strongly irradiated exoplanets. 
In addition to the usual difficulty of estimating the escape efficiency in
EUV-driven conditions,  
the proposed mechanism of Balmer-driven escape would require 
estimating how much of the NUV stellar energy is contributing to the escape, for
which there may not exist a simple answer.
\\

For comparison, we calculated $\dot{m}$
for the well-studied exoplanet HD209458b (Fig. \ref{mylifetime_fig}; cyan circles). 
The difference in $\dot{m}$ when considering the full (Sun-like) 
emission spectrum of its host star and when enforcing
$F^{\star}$($\lambda$$>$912 {\AA})$\equiv$0 is moderate. 
HD209458b's thermosphere does not build up abundant H(2) 
as to significantly tap into the comparatively weak stellar NUV emission. 
\\

\begin{figure}[h]
   \centering
   \includegraphics[width=8.cm]{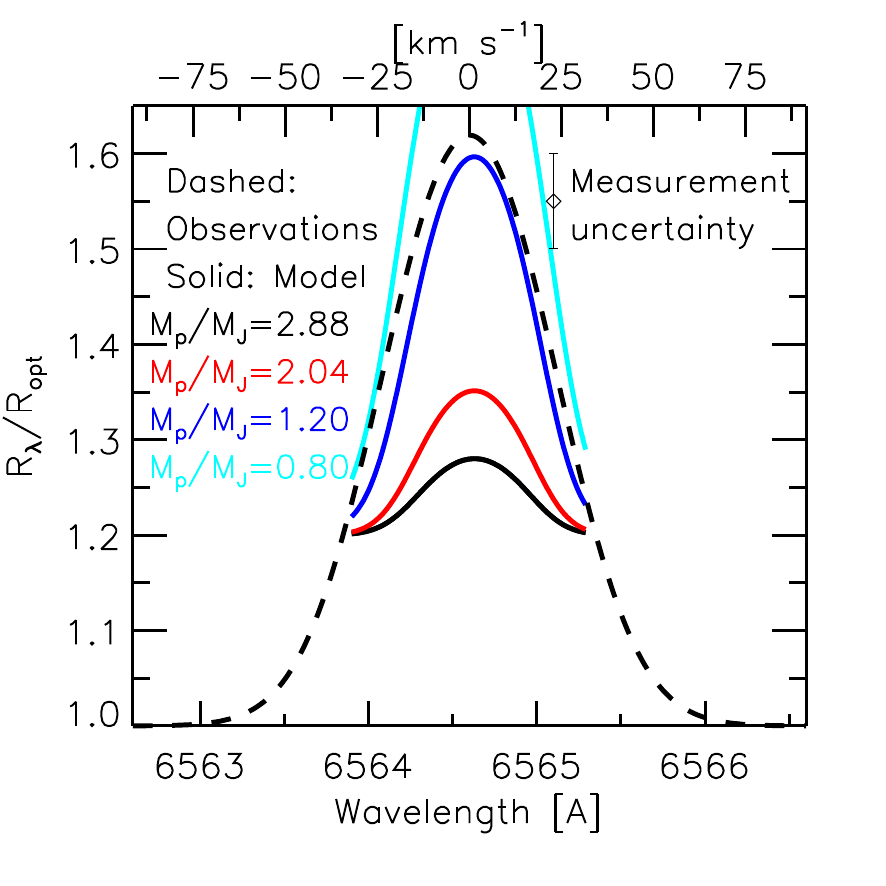}
   \caption{\label{myhaspectrum_fig} 
   H-$\alpha$ transmission spectrum. 
   Model-predicted effective size of KELT-9b at H-$\alpha$ 
   in units of optical radius ($R_{\rm{opt}}$/$R_{\rm{J}}$=1.9) for 
   $M_{\rm{p}}$/$M_{\rm{J}}$=2.88 (black), 2.04 (red), 
   1.20 (blue), and 0.80 (cyan). The synthetic spectrum considers the Doppler broadening
   introduced by the bulk atmospheric flow on both the day- and nightsides.   
   The dashed line represents the available H-$\alpha$ 
   measurements and their estimated uncertainties \citep{yanhenning2018, cauleyetal2019}.   
   }
   \end{figure}

It is instructive to compare our predicted transit depths 
in H-$\alpha$ with measurements \citep{yanhenning2018,cauleyetal2019}. 
For a meaningful comparison, it must be noted that the line core 
probes pressures within our thermospheric model domain, whereas
the line wing probes deeper down into the atmosphere 
up to the visible continuum level, 
which we estimate to occur at $p$$\sim$10$^{3}$ dyn cm$^{-2}$ (=1 mbar) 
\citep{lothringeretal2018}. From hydrostatic balance, 
we estimate the geometrical thickness of the region between
this and the $p$$\sim$0.1 dyn cm$^{-2}$ level as 
$\Delta$$r_{\rm{tr-reg}}$$\sim$4$H_{\rm{lt}}$$\ln{10}$, 
where $H_{\rm{tr-reg}}$ is an average scale height of this transition region.
Assuming that hydrogen is in atomic form 
\citep{kitzmannetal2018,lothringeretal2018}, a gravitational acceleration 
of 2,000 cm s$^{-2}$, and a characteristic temperature of 7,500 K 
in between KELT-9b's brightness temperature and our model predictions, 
we find that $H_{\rm{tr-reg}}$$\sim$3,000 km and 
$\Delta$$r_{\rm{tr-reg}}$$\sim$28,000 km or, equivalently, about 0.2$R_{\rm{p}}$. 
Using the H(2) predictions in our model at $p$$<$0.1 dyn cm$^{-2}$, we 
calculated the transmission spectrum in H-$\alpha$ and shifted it by 
$\Delta$$r_{\rm{tr-reg}}$ to approximately account 
for the thickness of the transition region above the visible continuum. 
The predictions match reasonably well the measured
transit depth for the smaller $M_{\rm{p}}$/$M_{\rm{J}}$ explored 
(Fig. \ref{myhaspectrum_fig}), 
which renders support to our proposed scenario of Balmer-driven escape: 
large H(2) column abundances imply substantial stellar NUV energy 
deposition and in turn rapid mass loss.
Taken at face value, the comparison suggests that KELT-9b's mass has been severely
overestimated or, alternatively, 
that our model predictions for H(2) along the substellar line are not fully
representative of the near-terminator conditions.
The imperfect match in the lineshape also suggests that other effects not 
considered in our substellar model 
such as wind motions 
\citep{tremblinchiang2013,trammelletal2014,shaikhislamovetal2018,
debrechtetal2019} play a role at shaping the flow probed during transit. 
Further modeling considering the three-dimensional geometry of the gas
will help elucidate the specifics of the near-terminator flow. 
We do not expect that this most welcome insight will modify the overall description of
the proposed Balmer-driven escape. 
\\

Our NLTE scheme is based on a hydrogen atom model. 
Interestingly, a few metals have been detected in the atmosphere 
of KELT-9b \citep{hoeijmakersetal2018,hoeijmakersetal2019,cauleyetal2019}, 
including the ion Fe$^{+}$ -- which is an efficient coolant. 
The altitude and abundance of these atoms remain poorly constrained though. 
It is worth assessing whether our findings are sensitive 
to moderate metallicity levels. 
To that end, we ran a few simulations in which we included a parameterization of
Fe$^{+}$ cooling \citep{gnatferland2012} with a density correction factor of $\sim$1/3 
\citep{wangetal2014} and up to solar Fe abundances. 
These simulations (not shown here) predict 
mass loss rates up to $\sim$1/3 smaller 
and thermospheric structures that are overall consistent with 
the simulations from the hydrogen-only model, 
but temperatures $\sim$1,000--2,000 K lower near the lower boundary. 
We did not explore higher metallicities, but this is a potentially interesting avenue 
for future work to set in context the existing measurements of
metals in KELT-9b's atmosphere.
\\

We presented a self-consistent model of hydrodynamics and NLTE in KELT-9b's
thermosphere. The model results show that the thermospheric energy budget of 
close-in planets irradiated by hot stars is dominated by H(2)
absorption of stellar NUV photons. This previously unrecognized source
of energy enhances the mass loss to possibly catastrophic rates. 

\newpage

\newpage

\newpage

\newpage

\acknowledgments

The authors acknowledge 
the support of the DFG priority program SPP 1992 
\textit{Exploring the Diversity of Extrasolar Planets} (Grant Numbers: GA 2557/1--1 
and SCHN 1382/2-1). PCS acknowledges funding support from DLR 50OR1901. 
The authors gratefully acknowledge correspondence with Jason Aufdenberg 
(Embry-Riddle Aeronautical University, USA) and Jorge Sanz-Forcada 
(Centro de Astrobiolog\'ia, CSIC-INTA, Spain). 
AGM thanks Juan Cabrera (DLR Adlershof, Germany) for a thorough reading of the manuscript. 
The authors gratefully acknowledge the insightful questions from the referees, 
which led to an improved manuscript.

\newpage

\appendix 

\setcounter{figure}{0}
\setcounter{table}{0}

We here describe the implementation of a NLTE scheme for
hydrogen atoms into a published hydrodynamic model.
\\

\section*{Hydrodynamic model} 

Our hydrodynamic model solves the mass, momentum and energy conservation equations 
in a planetary atmosphere  irradiated by its host star \citep{garciamunoz2007}. 
The model considers a gas that contains a total of $\mathcal{S}$=9 pseudo-species, and 
solves in a spherical geometry the conservation equations:
\begin{equation*}
\frac{\partial \mathbf{U} }{\partial t} + \frac{\partial \mathbf{F}}{\partial r} = \mathbf{S}
\label{structural1_eq}
\end{equation*}
\begin{equation*}
\mathbf{U} = \left(
\begin{array}{c}
 r^2 \rho_s  \\
 r^2 \rho u  \\
 r^2 \rho E \end{array} \right)
\qquad
\mathbf{F} = \left(
\begin{array}{c}
 r^2 \rho_s (u+u_s+v_s)  \\
 r^2 (\rho u^2+p ) \\
 r^2 ( \rho E +p)u + r^2 q \end{array} \right)
\label{structural2_eq}
\qquad
\mathbf{S} = \left(
\begin{array}{c}
 r^2 \dot{\omega}_s  \\
 r^2 \rho f_{\rm{ext}} + 2pr  \\
 \rho u r^2 f_{\rm{ext}} - r^2 \Gamma
\end{array} \right)
.
\end{equation*}
Here, $r$ is the radial distance to the planet center; 
$\rho_s$ is the mass density of the $s$th species, 
related to the number density and mass of the species through $\rho_s$=$n_s$$m_s$
and to the mass density of the whole gas through $\rho$=$\sum_{\mathcal{S}} \rho_s$; 
$u$ is the mass average velocity of the gas, and $u_s$ and $v_s$ are the molecular and eddy diffusion velocities of the 
$s$th species, respectively;
$p$ is pressure; 
$\rho E = \rho e + \frac{1}{2} \rho u^2 $ is the total energy of the gas, 
where the internal energy 
$\rho e$=$\sum_{\mathcal{S}} n_s e_s^{\rm{tran}}$+$\sum_{\mathcal{S}-e} n_s E^{\rm{exci}}_s $ 
includes contributions from 
the translational motion of all species ($e_s^{\rm{tran}}$=$\frac{3}{2}kT$)
and the excitation energy above the ground state for the atoms ($E^{\rm{exci}}_s$); 
$\dot{\omega}_s$ is the net mass production for the $s$th species; 
$f_{\rm{ext}}$ is the net external force per unit of volume, which includes 
contributions from the planet and stellar gravitational attractions and 
from the planet's centrifugal motion; $q$ is the heat flux 
(with contributions from thermal conduction and the transport of 
enthalpy by diffusion of each species); 
$\Gamma$ is the net energy emission rate from radiative processes.
\\

We discretize the above conservation equations over a spatial grid from 1 to 2.5 
in $r$/$R_p$. It comprises 260 cells of increasing size, from $\sim$30 km
at the lower boundary to $\sim$3,000 km near the top. 
The solution near the lower boundary is strongly influenced by heating from above, which
results in temperatures near $p$$\sim$0.1 dyn cm$^{-2}$ of 
$\sim$10,000 K, more than twice KELT-9b's equilibrium temperature. 
Our choice of a bottom boundary temperature of 8,000 K 
aims at partly minimizing the gradient in temperature and velocity 
near the lower boundary seen in Fig. (\ref{mypanel_fig}).
Imposing alternative temperatures of 4,600 or 10,000 K at the bottom boundary 
affects the calculated mass loss rates by less than 2\%. In other words, 
the choice of temperature there is not critical for the overall solution.
\\

In the simulations presented here  
we generally placed the lower boundary of the model at a pressure of 0.1 dyn cm$^{-2}$. 
We did also explore the effect of shifting the lower boundary to 0.2 dyn cm$^{-2}$. 
For consistency, in this simulation we 
reduced the 
radial distance of the lower boundary to 1.83$R_{\rm{J}}$, which  
ensures that at $r$$\sim$1.89$R_{\rm{J}}$ the pressure
remains $\sim$0.1 dyn cm$^{-2}$. The new simulation resulted in a mass loss rate 
$\sim$3\% larger than in the standard setting. 
This small difference suggests that heating at even higher pressures 
will have a minor impact on the mass loss rate, and that for it to drive the escape
the energy must be deposited at high enough altitude.
\\

For the calculations, we adopted a time step of 0.2 seconds to ensure stability of the
time marching scheme. This is a few times less than for similar calculations that we did 
without the NLTE scheme. 
We interpret this more stringent requirement on the time step as a 
consequence of the strong interaction between the hydrodynamic and radiative problems.
As usual, it is convenient to initialize a calculation with a previously converged solution
and gradually modify the input conditions. 
Solving the radiative transfer equation in the NLTE scheme takes a significant part of the 
total computational time, which can be of a few days on a single CPU 
for full convergence.
To speed up the calculations, it is convenient to start with a relatively coarse
spectral grid and switch later to the finer spectral grid.

\section*{NLTE scheme}
Table (\ref{hatom_table}) summarizes the adopted hydrogen atom model, where
index $i$=1$-$7 simply specifies each bound state.
We treat each bound state, the protons and the electrons as separate
pseudo-species connected with one another through collisional and radiative processes. 
This treatment builds upon a long history of NLTE schemes for plasmas \citep{batesetal1962a,batesetal1962b}, 
and more particularly on the scheme for stellar atmospheres of \cite{munafoetal2017}. 
The reader is referred to \cite{munafoetal2017} for a thorough description of the fundamental
collisional and radiative processes. 
The work of \cite{munafoetal2017} assumes
that the pseudo-species in the plasma follow Maxwellian distributions of velocities at two 
specified translational temperatures: 
$T_{\rm{e}}$ for electrons and $T_{\rm{h}}$ for heavy particles. 
Unlike in that work, our scheme assumes that
a single temperature $T$=$T_{\rm{e}}$=$T_{\rm{h}}$ describes the velocities of both electrons and heavy particles, 
which simplifies the treatment of the energy balance.
Future work should look into the impact of this simplification.

\begin{table}[h]
\caption{Hydrogen atom model. Information from the NIST Bibliographic Database \citep{kramida2010}.}             
\label{hatom_table}      
\centering               
\begin{tabular}{c c c c} 
\hline\hline             
Index & Denomination & Degeneracy  & Energy \\  
 $i$ &  &  $g_i$ & $E^{\rm{exci}}_i$ [eV] \\    
\hline                        
1 & $1s$ $^2S_{1/2}$    & 2 & 0 \\
2 & $2p$ $^2P^*_{1/2}$  & 2 & 10.1988061 \\
3 & $2s$ $^2S_{1/2}$    & 2 & 10.1988104 \\
4 & $2p$ $^2P^*_{3/2}$  & 4 & 10.1988514 \\
5 & $3$ & 18 & 12.0875051 \\
6 & $4$ & 32 & 12.7485392 \\
7 & $5$ & 50 & 13.0545016 \\
$\infty$ & Continuum, H$^{+}$ & $g_{\infty}$=1  & $E_{\infty}$=13.5984345 \\
$e$ & Electron & $g_{e}$=2  &  \\
\hline \hline
\hline                                   
\end{tabular}
\end{table}

\subsection*{Collisional processes}
We considered 
excitation ($i$$\rightarrow$$j$) and de-excitation ($j$$\rightarrow$$i$)
processes\footnote{We generally follow
the convention $E_i$$<$$E_j$ for transitions between bound states with indices $i$ and $j$.}
 in collisions of bound states with electrons:
\begin{equation*}
\mbox{H}(i) + \mbox{e} \xrightleftharpoons[k^e_{j \rightarrow i}\;{[\rm{cm}^3\rm{s}^{-1}]} ]
{k^e_{i \rightarrow j}\;{[\rm{cm}^3\rm{s}^{-1}]}} \mbox{H}(j) + \mbox{e}, 
\label{collexcitation_eq}
\end{equation*}
as well as ionization ($i$$\rightarrow$$\infty$) processes and 
their reverse of three-body recombination ($\infty$$\rightarrow$$i$): 
\begin{equation*}
\mbox{H}(i) + \mbox{e} \xrightleftharpoons[k^e_{\infty \rightarrow i}\;{[\rm{cm}^6\rm{s}^{-1}]}]{k^e_{i \rightarrow \infty}
\;{[\rm{cm}^3\rm{s}^{-1}]}} \mbox{H}^+ + \mbox{e} + \mbox{e}. 
\label{collionization_eq}
\end{equation*}
Rate coefficients were collected from a variety of sources. 
For excitation
\citep{andersonetal2000,andersonetal2002,przybillabutler2004}:
\begin{equation*}
k^e_{i\rightarrow j} = 
\frac{8.63\times10^{-6}}{g_i T^{1/2}} \Upsilon_{ij} 
\exp{\Big\{-\frac{E_j-E_i}{kT_e}} \Big\}, 
\end{equation*}
where $\Upsilon_{ij}$ is the Maxwell-averaged effective collision strength. 
$\Upsilon_{ij}$ varies moderately with temperature and for simplicity we took
the values specific to $\sim$10$^4$ K.
From detailed balance, the de-excitation rate coefficient is:
\begin{equation*}
k^e_{j\rightarrow i} = 
\frac{8.63\times10^{-6}}{g_j T^{1/2}} \Upsilon_{ij}. 
\end{equation*}
We adopted $\Upsilon_{ij}$ from
\cite{andersonetal2002} when the orbital quantum number of the
bound states are specified, and from \cite{przybillabutler2004} otherwise. 
For collisions between degenerate states
$2p$ $^2P^*_{1/2}$$-$$2s$ $^2S_{1/2}$ and $2s$ $^2S_{1/2}$$-$$2p$ $^2P^*_{3/2}$, 
we adopted the $\Upsilon_{ij}$ reported by \cite{aggarwaletal2018}. 
\\

For ionization 
we adopted the formulae published by \cite{barklem2007} for states with principal quantum number
$n$$\le$2 and those by \cite{vrienssmeets1980} for $n$$>$2. 
We re-fitted the published formulae
to expressions of the form $a T^{b} \exp{(c/T)}$ for temperatures between
4$\times$10$^3$ and $1.4$$\times$10$^4$ K, and implemented the latter. 
Our fits are accurate to within 20\% 
with respect to the original formulae over the quoted temperatures.
Table (\ref{e_ionization_table}) summarizes the implemented rate coefficients for
electron-collision ionization $k^e_{i\rightarrow \infty}$. 
The rate coefficients for three-body recombination 
${k^e_{\infty \rightarrow i}}$ 
are related to the rate coefficients for ionization through:
\begin{equation*}
\frac{k^e_{\infty \rightarrow i}} {k^e_{i \rightarrow \infty}}=\frac{g_i}{g_e g_{\infty}}
\frac{h^3}{(2\pi m_e k T_e)^{3/2}} \exp{\Big\{\frac{E_{\infty}-E_i}{kT_e}\Big\}},
\end{equation*}
where the degeneracies $g_e$=2 and $g_{\infty}$=1, $h$ and $k$ are the Planck and
Boltzmann constants, and $m_e$ is the electron mass.
\\

\begin{table}[h]
\caption{Rate coefficients for electron-collision ionization. 
For each entry, the three numerical values specify, from top to bottom: $a$, $b$, $c$, with
the rate coefficient $a T^{b} \exp{(c/T)}$ [cm$^3$ s$^{-1}$].
Numbers in parentheses stand for exponents, i.e. 1.35($-$7) means 1.35$\times$10$^{-7}$. 
}             
\label{e_ionization_table}
\centering                          
\begin{tabular}{c c c}        
\hline\hline                 
Index &  $k^e_{i\rightarrow \infty}$ & Reference \\    
 $i$ &  [cm$^3$ s$^{-1}$]  &   \\    
\hline                        
 &   1.51 &\\
1 &  $-$2.09  & \cite{barklem2007} \\
 &    $-$1.68(+5) &  \\
 \hline
 &  1.35($-$7) & \\
2--4 &  $-$0.033 & \cite{barklem2007} \\
 &  $-$3.89(+4) & \\
 \hline 
 &  1.96($-$7) &  \\
5 & 0.15 & \cite{vrienssmeets1980}  \\
 &  $-$1.85(+4) & \\
 \hline
 &  1.27($-$6) &  \\
6 & 0.053 &  \cite{vrienssmeets1980} \\
 &  $-$1.09(+4) & \\
 \hline
 &  5.39($-$6) & \\
7 &  $-$0.023  & \cite{vrienssmeets1980} \\
 &   $-$7.38(+3) & \\
 \hline
\end{tabular}
\end{table}

The mixing of nearly-equal energy states in collisions with protons is very rapid. 
Thus, we considered: 
\begin{equation*}
\mbox{H}(i) + \mbox{H}^+ \xrightleftharpoons[k^p_{j \rightarrow i}\;{[\rm{cm}^3\rm{s}^{-1}]}]
{k^p_{i \rightarrow j}\;{[\rm{cm}^3\rm{s}^{-1}]}} \mbox{H}(j) + \mbox{H}^+, 
\label{prontonexcitation_eq}
\end{equation*}
for the $2p$ $^2P^*_{1/2}$, $2s$ $^2S_{1/2}$ and $2p$ $^2P^*_{3/2}$ states. 
We adopted the rate coefficients at $T$=10$^4$ K calculated by \cite{seaton1955},
omitting the mild temperature-dependence of the process \citep{struenseecohen1988}.
Neglecting the inter-state energy separation with respect to $kT$, 
the reverse rates were simply estimated from the ratios of degeneracies.
\\

The processes involving collisions with neutrals: 
\begin{eqnarray*}
\mbox{H}(i) + \mbox{H}(n=1) \rightleftharpoons \mbox{H}(j) + \mbox{H}(n=1) \\
\mbox{H}(i) + \mbox{H}(n=1) \rightleftharpoons \mbox{H}^+ + \mbox{e} + \mbox{H}(n=1)
\label{hi_h1_ion_eq}
\end{eqnarray*}
are outpaced by their electron-collision counterparts even for small 
ionization fractions of $\sim$10$^{-4}$
\citep{colonnaetal2012,munafoetal2017}. 
These processes are safely neglected for the conditions of
KELT-9b's thermosphere. 
\\

Collisional processes contribute to the population of 
bound states,  protons and electrons through the corresponding 
mass production terms $\dot{\omega}_i$, $\dot{\omega}_{\rm{H}^+}$, and $\dot{\omega}_e$ 
in the gas continuity equations.

\subsection*{Radiative processes}

Our NLTE scheme considers bound-bound (BB), bound-free/free-bound (BF/FB) and free-free (FF)
radiative transitions. 
These processes couple the radiation field with the population of hydrogen atom states, 
and introduce non-local effects as photons 
diffuse through the atmosphere from where they are emitted to where 
they are ultimately absorbed.
In our treatment, we omitted the process of induced emission in BB and BF/FB transitions
because its effect is minor.

\subsubsection*{BB transitions}

We considered photoexcitation between bound states by absorption:
\begin{equation*}
\mbox{H}(i) + hc/\lambda_{ij} \xrightarrow[]{(B_{ij}\lambda^2_{ij}/c ) \Phi_{ij} \;{[\rm{s}^{-1}]}} \mbox{H}(j) 
\end{equation*}
and spontaneous emission: 
\begin{equation*}
\mbox{H}(j) \xrightarrow[]{A_{ji}\;{[\rm{s}^{-1}]}} \mbox{H}(i) + hc/\lambda_{ij}. 
\end{equation*}
$c$ is the speed of light and $\lambda_{ij}$=$hc$/($E_j$$-$$E_i$) 
the line wavelength. 
$B_{ij}$ and $A_{ji}$ are the Einstein coefficients for absorption and 
spontaneous emission, related through:
\begin{equation*}
B_{ij}=\frac{g_j}{g_i}\frac{\lambda^3_{ij}}{2hc} A_{ji}.
\end{equation*}
$\Phi_{ij}$=$\int \mathcal{J}_{\lambda} \psi^{ij}_{\lambda} d\lambda $ [erg\;cm$^{-2}$s$^{-1}$sr$^{-1}$cm$^{-1}$]
is the integral of the local 
average intensity $\mathcal{J}_{\lambda}$ (see below)
over the transition line profile $\psi^{ij}_{\lambda}(\lambda)$
(normalized to $\int_{0}^{\infty}\psi^{ij}_{\lambda}(\lambda) d\lambda$=1).
We assume complete frequency redistribution, and thus the line profiles 
for absorption and emission satisfy $\psi^{ij}_{\lambda}(\lambda)=\psi^{ji}_{\lambda}(\lambda)$.
We model $\psi^{ij}_{\lambda}(\lambda)$ as a Voigt function obtained
from the convolution of Gaussian and Lorentz functions, and
use the analytical approximations given by \cite{whiting1968}.
Thermal broadening contributes to the Gaussian component, and natural 
and Stark broadening contribute to the Lorentz component.
Our implementation of Stark broadening is based on low pressure plasmas
\citep{griem1974,kramidaetal2018}. 
For computational expediency, we considered for photoexcitation and in the radiative transfer 
solution only the 
{\lalpha} line that pumps ground state atoms into the excited states
$2p$ $^2P^*_{1/2}$ and $2p$ $^2P^*_{3/2}$.
This approximation is adequate to describe the population of excited states with 
principal quantum number $n$=2, which is the focus of our work. 
All the other lines are not considered for photoexcitation or in the radiative transfer
problem. We do consider however the effect of spontaneous emission of all 
BB transitions in the energy balance
by considering that the emission lines other than {\lalpha} are optically transparent 
and thus the corresponding radiated energy escapes the atmosphere.
That absorption in {\lalpha} is more important than in the other
-$\beta$, -$\gamma$, -$\delta$, etc., lines of the Balmer series is justified by the combination
of decreasing values for the corresponding 
$B_{1j}$ Einstein coefficients and the drop of stellar emission towards
shorter wavelengths.
\\
 
We expect that a more complete treatment of all BB transitions will increase to some extent 
the population of $n$=2 states, as the states with $n$$>$2 will cascade
through $n$=2. We also expect that solving the radiative transfer problem in all the lines, 
i.e. moving away from the assumption of transparency for all lines except {\lalpha}, 
will increase the energy deposited in the thermosphere. 
Either way, these changes -- which will be implemented in future work -- will
reinforce the mechanism that sustains the proposed Balmer-driven escape. 
\\

For BB transitions, the emission
$\varepsilon^{\rm{BB}}_{\lambda}$ [erg\;cm$^{-3}$s$^{-1}$sr$^{-1}$cm$^{-1}$]
and absorption $\kappa^{\rm{BB}}_{\lambda}$ [cm$^{-1}$]
coefficients that enter the radiative transfer equation are generally given by:
\begin{equation*}
\varepsilon^{\rm{BB}}_{\lambda}=\sum_{j>i} \frac{hc}{4\pi\lambda_{ij}} 
n_j A_{ji} \psi^{ij}_{\lambda} 
\end{equation*}
\begin{equation*}
\kappa^{\rm{BB}}_{\lambda}=\sum_{j>i} \frac{h \lambda_{ij}}{4\pi} n_i B_{ij}   \psi^{ij}_{\lambda}, 
\end{equation*}
where $n_i$ is the number density of bound state $i$, and 
$\kappa^{\rm{BB}}_{\lambda}$ omits the typically small contribution from induced emission. 
In our treatment, we use the above expressions for $\varepsilon^{\rm{BB}}_{\lambda}$ and
$\kappa^{\rm{BB}}_{\lambda}$ only for {\lalpha}. For the other lines, we 
take $\kappa^{\rm{BB}}_{\lambda}$$\approx$0 and simply include the wavelength-integrated form of
$\varepsilon^{\rm{BB}}_{\lambda}$ in the net energy emission rate $\Gamma$.
\\
 
Our adopted Einstein coefficients are based on transition probabilities and wavelengths 
from the NIST Bibliographic Database \citep{kramidaetal2018}.

\subsubsection*{BF/FB transitions}

We considered photoionization of bound states:
\begin{equation*}
\mbox{H}(i) + hc/\lambda (>E_{\infty}-E_i) 
\xrightarrow[]{J_{i \rightarrow \infty} \;{[\rm{s}^{-1}]} }
\mbox{H}^+ + \mbox{e} 
\end{equation*}
and spontaneous radiative recombination:
\begin{equation*}
\mbox{H}^+ + \mbox{e}
\xrightarrow[]{k^{rr}_{\infty \rightarrow i} \;{[\rm{cm}^3\rm{s}^{-1}]} }
\mbox{H}(i) + hc/\lambda (>E_{\infty}-E_i).
\end{equation*}

For BF/FB transitions, we calculated the emission and absorption coefficients from:
\begin{equation*}
\varepsilon^{\rm{FB}}_{\lambda}=\frac{h^4 c^2 n_e n_{\rm{H}^+}}{\lambda^5 (2\pi m_e k T_e)^{3/2}} \times
\sum_i \sigma_i^{\rm{PI}} (\lambda) \frac{g_i}{g_{\infty}} 
\exp{\Big\{\frac{E_{\infty}-E_i}{kT_e}}  - \frac{hc/\lambda}{kT_e} \Big\}
\end{equation*}
\begin{equation*}
\kappa^{\rm{BF}}_{\lambda}=
\sum_i \sigma_i^{\rm{PI}} (\lambda) n_i.
\end{equation*}
$\kappa^{\rm{BF}}_{\lambda}$ neglects the small contribution from induced radiative recombination. 
We calculated the photoionization cross sections 
$\sigma_i^{\rm{PI}} (\lambda)$ with the analytical formula for hydrogenic atoms given by
\cite{mihalas1978} (pp. 99, Eq. 4-114), 
and omitted the small corrections from the bound-free Gaunt factors. 
\\

\begin{figure*}
   \centering
   \includegraphics[width=9.cm]{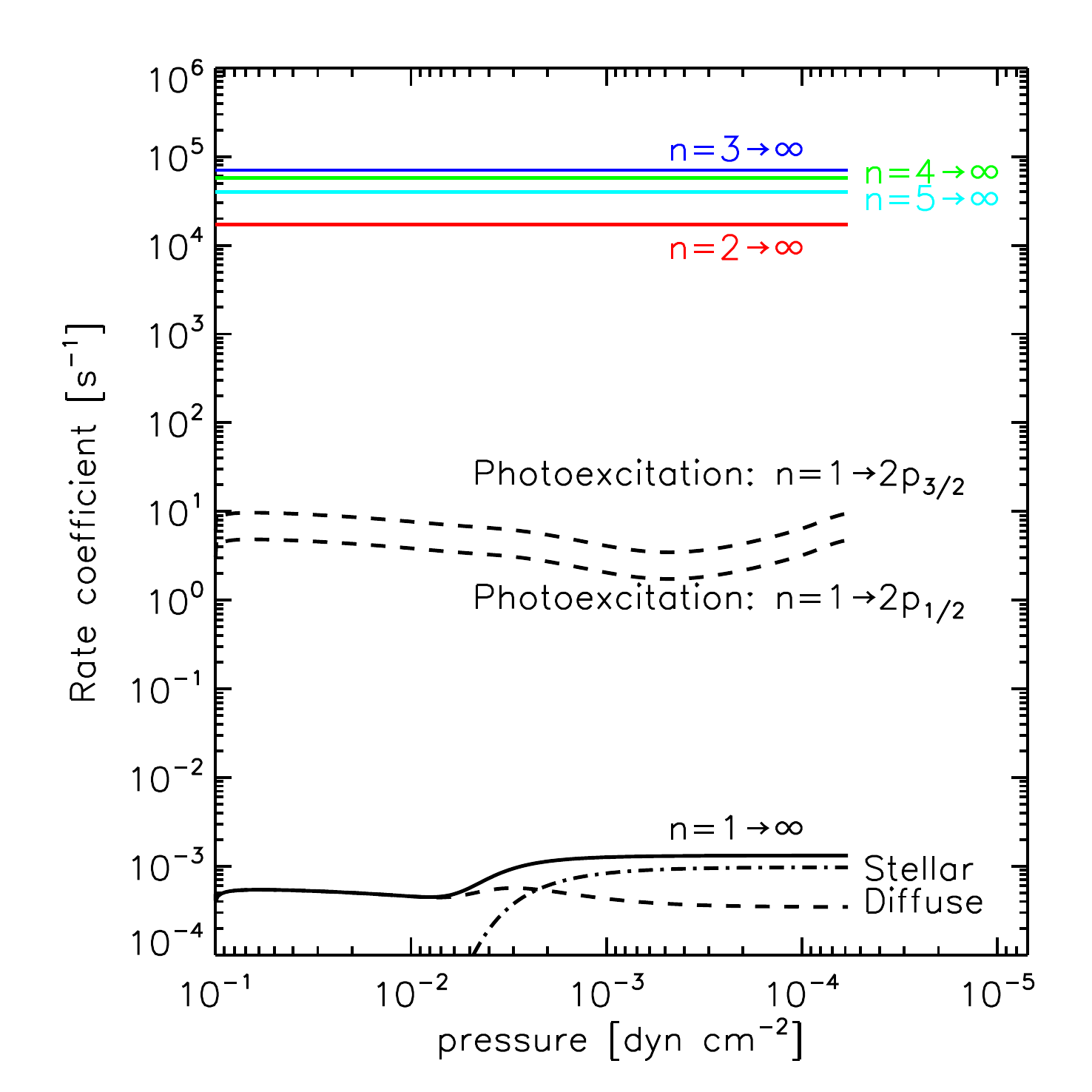}
      \caption{\label{kf_fig} Photoionization and photoexcitation rate coefficients.
      For photoionization of H(1), the dotted-dashed and dashed lines distinguish the
      separate contributions from stellar and diffuse radiation.
      }
   \end{figure*}

The rate coefficient for photoionizaton is calculated as: 
\begin{equation*}
J_{i \rightarrow \infty}=\frac{4\pi}{hc} \int \sigma_i^{\rm{PI}}(\lambda)  \mathcal{J}_{\lambda}(\lambda) \lambda d\lambda,
\end{equation*}
and for spontaneous radiative recombination:
\begin{equation*}
k^{rr}_{\infty \rightarrow i}=\sqrt{\frac{2}{\pi}} \frac{g_i}{g_{\infty}} \frac{h^3 c}{(m_e k T_e)^{3/2}} 
 \exp{\Big\{\frac{E_{\infty}-E_i}{kT_e}} \Big\}
 \int \frac{\sigma_i^{\rm{PI}}(\lambda)}{\lambda^4}  \exp{\Big\{-\frac{hc/\lambda}{kT_e}} \Big\} d\lambda.
\end{equation*}
We integrated numerically $k^{rr}_{\infty \rightarrow i}$ 
and fitted the resulting rate coefficients at temperatures of 10$^3$--2$\times$10$^4$ K 
to the expression $a T^{b} \exp{(c/T)}$. 
Radiative recombination readily populates highly excited states.  
Because our atom model is truncated at the principal quantum number $n$=5, 
we calculated the 
rate coefficients up to $n$$\sim$40 and
added the net difference for channels above $n$=4 to our channel with $n$=5, thus
ensuring that the net recombination of protons and electrons occurs at the 
proper rate. 
Table (\ref{JphotoionizationTOA_table}) lists the
photoionization rate coefficients at the top of the atmosphere 
for our choice of stellar spectrum.
Figure (\ref{kf_fig}) shows the variation of the photoionization rate coefficients 
with altitude, together with the rate coefficients 
$(B_{ij}\lambda^2_{ij}/c) \Phi_{ij}$ for photoexcitation 
$1s$ $^2S_{1/2}$$\rightarrow$$2p$ $^2P^*_{1/2}$ and 
$1s$ $^2S_{1/2}$$\rightarrow$$2p$ $^2P^*_{3/2}$.

\begin{table}[h]
\caption{Photoionization rate coefficient at the top of the atmosphere.}             
\label{JphotoionizationTOA_table}      
\centering                          
\begin{tabular}{c c }        
\hline                 
Principal quantum number & \multicolumn{1}{c}{${J_{n \rightarrow \infty} }$ } \\
 $n$ & [s$^{-1}$] \\ 
\hline 
1    & 1.34($-$3) \\
2    & 1.70($+$4) \\
3    & 7.03($+$4) \\
4    & 5.73($+$4) \\
5    & 3.99($+$4) \\
\hline
\end{tabular}
\end{table} 

\subsubsection*{FF transitions}

For FF transitions the emission and absorption coefficients are:
\begin{equation*}
\varepsilon^{\rm{FF}}_{\lambda}=\frac{8}{3} \sqrt{\frac{2\pi}{3m_ekT_e}} \frac{n_e n_{\rm{H}^+} q^6_{\rm{e}}} {\lambda^2 m_e c^2}
 \exp{\Big\{-\frac{hc/\lambda}{kT_e}} \Big\}
\end{equation*}
\begin{equation*}
\kappa^{\rm{FF}}_{\lambda}=\frac{4}{3} \sqrt{\frac{2\pi}{3m_ekT_e}} \frac{\lambda^3 n_e n_{\rm{H}^+} q^6_{\rm{e}}} {m_e h c^4}
\Big[ 1- \exp{\Big\{-\frac{hc/\lambda}{kT_e}} \Big\} \Big]
\end{equation*}
where $q_e$=4.8032068$\times$10$^{-10}$	esu is the electron charge in cgs units.
\\

\subsubsection*{Contribution to populations }
BB and BF/FB transitions contribute to the population 
of bound states, protons and electrons through the corresponding 
mass production terms $\dot{\omega}_i$, $\dot{\omega}_{\rm{H}^+}$, and $\dot{\omega}_e$.
\\

\section*{Radiative transfer}\label{subsec:radiativetransfer}

We solve the radiative transfer equation:
\begin{equation*}
\frac{dI_{\lambda}}{ds}=-\kappa_{\lambda}I_{\lambda} + \varepsilon_{\lambda}, 
\end{equation*}
which does not explicitly include a scattering term, 
although scattering occurs both in BB and BF/FB transitions. 
For instance at {\lalpha} a photon emitted in the optically thick line will be reabsorbed 
immediately after, which will lead to a new reemission.
Also, absorption of energetic photons will ionize the neutral gas, 
the products of which will recombine and reemit a fraction of the initially absorbed radiation. 
Each new absorption-emission event is effectively a scattering event. 
The above equation also does not explicitly consider the redistribution of photons in 
wavelength by e.g. the Doppler shift induced by
the thermal motion of atoms or the bulk gas motion. 
Although these effects will modify aspects of the radiation field such as
the penetration of photons in the thermosphere \citep{huangetal2017}, 
their rigorous treatment in the framework of our hydrodynamics-NLTE model 
is currently impractical. 
\\

\begin{figure*}
   \centering
   \includegraphics[width=9.cm]{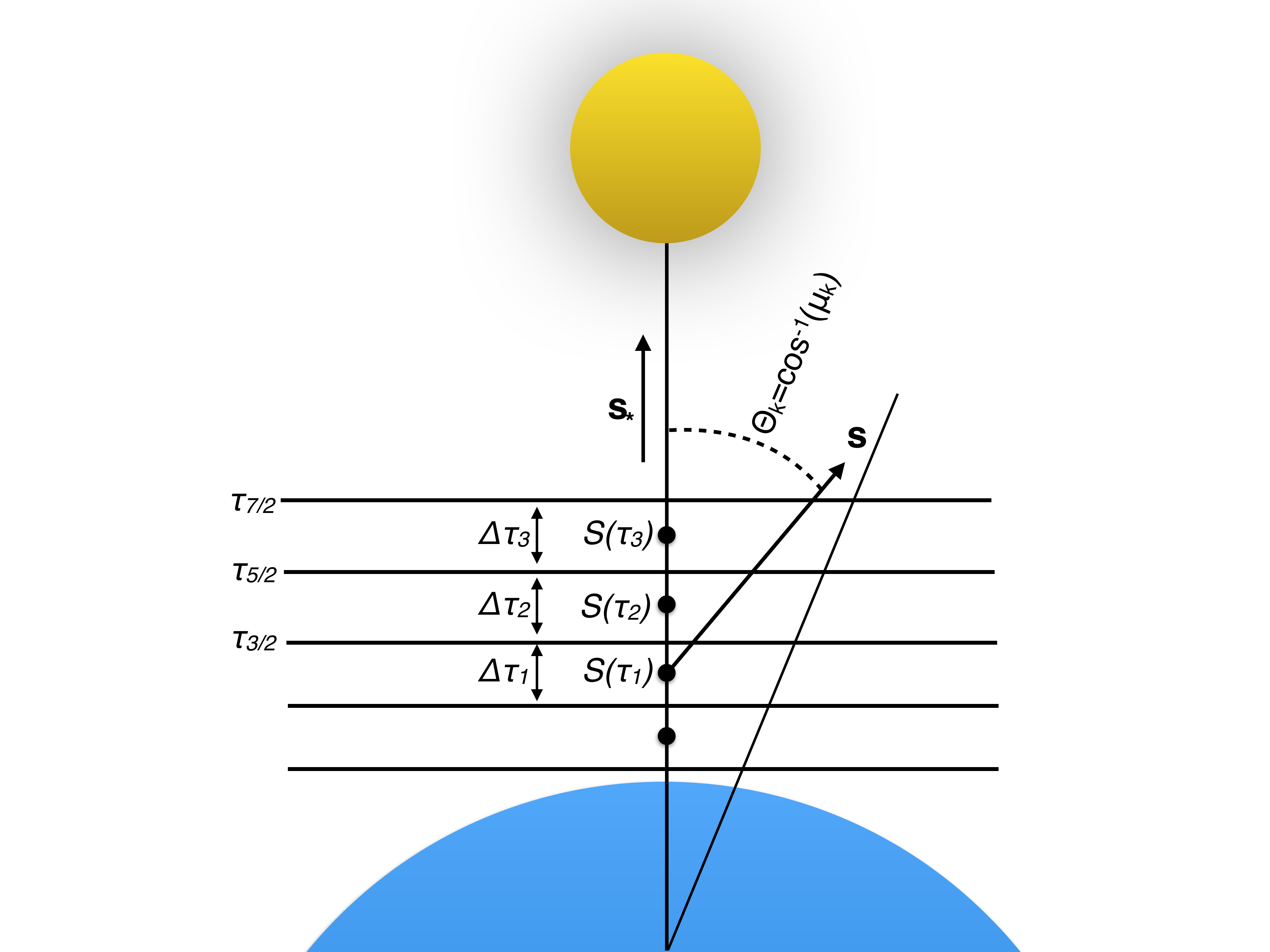}
      \caption{\label{rte_kelt_sketch_fig} 
  The radiative transfer equation is solved in the approximation of plane-parallel 
  atmosphere. The diffuse radiances $I^{\rm{d}}_{\lambda}(\mathbf{x,\mu}_{k})$ are solved 
  along multiple directions specified by the angle $\Theta_k$ and 
  integrated in solid angle to obtain the diffuse average intensity $\mathcal{J}^{\rm{d}}_{\lambda}$.
   }
   \end{figure*}

We solve the radiative transfer equation  
in a plane-parallel atmosphere along the substellar line
(Fig. \ref{rte_kelt_sketch_fig}), obtaining the
radiance $I_{\lambda}$$(\mathbf{x,s})$ [erg\;cm$^{-2}$s$^{-1}$sr$^{-1}$cm$^{-1}$].
The solution to the radiative transfer equation at a location $\mathbf{x}$ for radiation coming 
from direction $-\mathbf{s}$ is:
\begin{equation}
I_{\lambda}(\mathbf{x,s})=
 I_{\lambda}^{\star}(\mathbf{x,s}) +
 I_{\lambda}^{\rm{d}}(\mathbf{x,s})=
\mathcal{F}^{\star}_{\lambda}
\exp{(-\tau_{\lambda}(\mathbf{x}\rightarrow\mathbf{x}_{\star}))}
 \delta(\mathbf{s}-\mathbf{s}_{\star})  
 +
\int_{\mathbf{x}}^{\infty} S_{\lambda} (\mathbf{x'})
\exp{(-\tau_{\lambda}(\mathbf{x}\rightarrow\mathbf{x'}))}
d\tau_{\lambda}(\mathbf{x'}). 
 \label{ilambda_eq}
\end{equation}
$I_{\lambda}^{\star}(\mathbf{x,s})$ represents the non-diffuse radiance associated with the
stellar irradiance $\mathcal{F}^{\star}_{\lambda}$ [erg cm$^{-2}$ s$^{-1}$ cm$^{-1}$] 
at the planet's orbit that is attenuated through the atmosphere. 
$I_{\lambda}^{\star}(\mathbf{x,s})$ is zero for all directions except for the direction 
towards the star $\mathbf{s}_{\star}$.
$\delta(\mathbf{s}-\mathbf{s}_{\star})$ is the Dirac delta function centered at $\mathbf{s}_{\star}$
and by definition $\int  \delta(\mathbf{s}-\mathbf{s}_{\star}) d\Omega(\mathbf{s})$=1. 
$I_{\lambda}^{\rm{d}}(\mathbf{x,s})$ represents the diffuse radiance that arises within the atmosphere due to 
BB, FB and FF emission, $d\tau_{\lambda}(\mathbf{x'})$=$\kappa_{\lambda}(\mathbf{x'}) d\ell$
and $\ell$=$\|\mathbf{x}-\mathbf{x'}\|$. 
$S_{\lambda} (\mathbf{x})$=$\varepsilon_{\lambda} (\mathbf{x})$/$\kappa_{\lambda}(\mathbf{x})$
is the source function.
One could conceive an additional contribution to $I_{\lambda}(\mathbf{x,s})$ from 
radiation originating below the thermosphere, but we confirmed that this contribution
is negligible.
\\

\begin{figure*}
   \centering
   \includegraphics[width=20.cm, angle=+90]{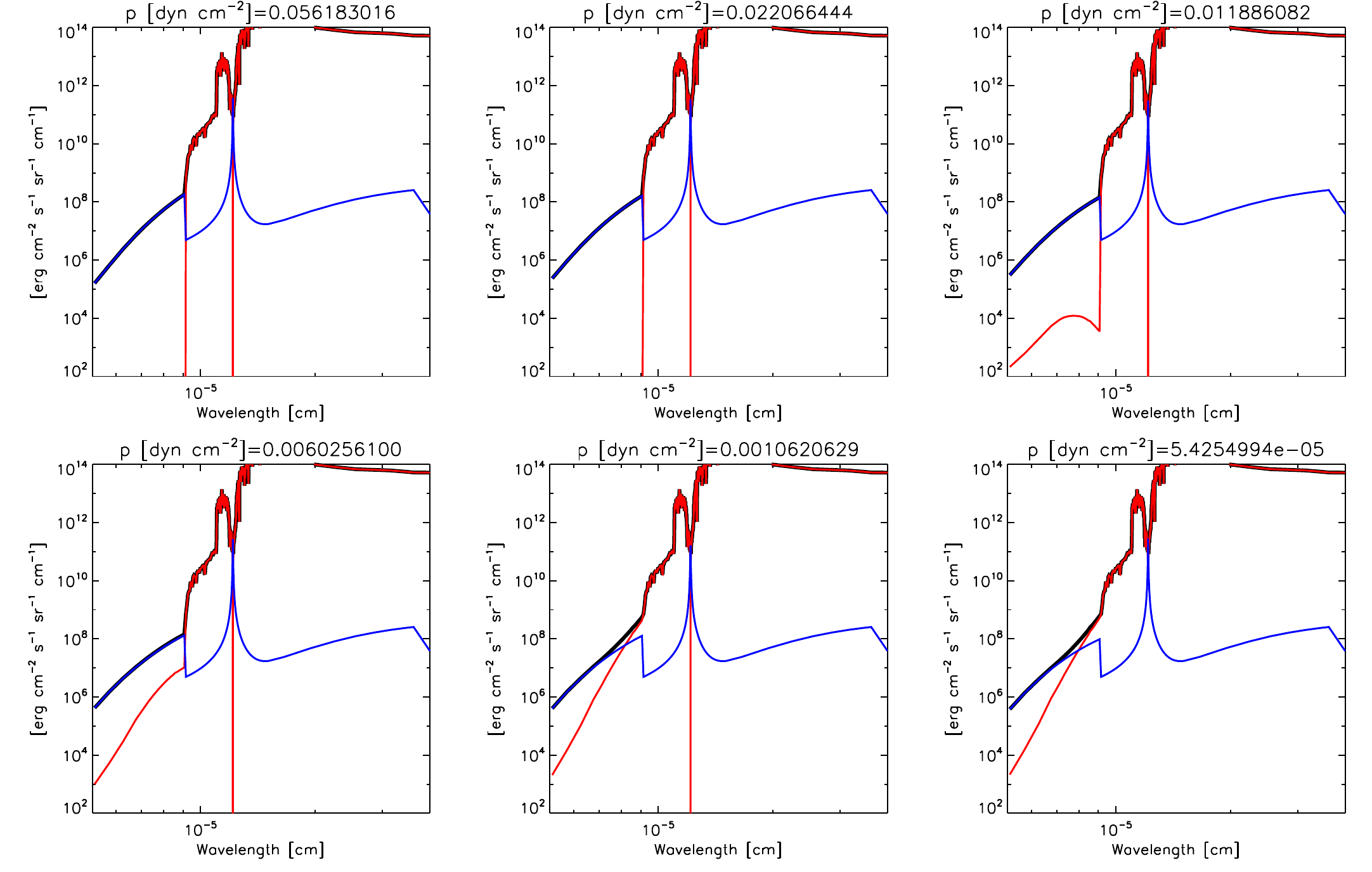}
      \caption{\label{plot_averdn_fig} 
      For the fiducial model, average intensity ($\mathcal{J}_{\lambda}$; black) and 
      contributions from the star ($\mathcal{J}^{\star}_{\lambda}$; red) and from
      diffuse radiation ($\mathcal{J}^{\rm{d}}_{\lambda}$; blue) at the specified pressure levels.      
      }
   \end{figure*}

The average intensity 
$\mathcal{J}_{\lambda}(\mathbf{x})$ (same units as $I_{\lambda}(\mathbf{x,s})$) that
contributes to various BB and BF/FB processes is defined as the integral of radiance over solid angle 
$\Omega(\mathbf{s})$:
\begin{equation*}
\mathcal{J}_{\lambda}(\mathbf{x})=\frac{1}{4\pi} \int I_{\lambda}(\mathbf{x}, \mathbf{s}) d\Omega(\mathbf{s}).
\end{equation*}
For the stellar contribution:
\begin{equation*}
\mathcal{J}^{\star}_{\lambda}(\mathbf{x})=\frac{\mathcal{F}^{\star}_{\lambda}}{4\pi}
 \exp{(-\tau_{\lambda}(\mathbf{x}\rightarrow\mathbf{x}_{\star}))}. 
\end{equation*}
The expression for $I_{\lambda}^{\rm{d}}(\mathbf{x,s})$ 
 is a line integral that can be evaluated once the gas properties are specified. 
To capture the directionality of diffuse radiation, $I_{\lambda}^{\rm{d}}(\mathbf{x,s})$
is calculated at $n_{\mu}$ evenly-separated directions 
$\mu_k$=$\mathbf{s}_k\cdot\mathbf{u_r}$=$\cos{\Theta_k}$=$-1$+2/$n_{\mu}$$\times$($k$$-$1/2)
and $k$=1, ..., $n_{\mu}$. 
We numerically calculate the diffuse average intensity as:
\begin{equation*}
\mathcal{J}^{\rm{d}}_{\lambda}(\mathbf{x})=\frac{1}{4\pi} \int I^{\rm{d}}_{\lambda}(\mathbf{x}, \mathbf{s}) d\Omega(\mathbf{s})
=\frac{1}{n_{\mu}}\sum_{k=1}^{n_{\mu}} I^{\rm{d}}_{\lambda}(\mathbf{x,\mu}_{k}).
\end{equation*}
We take $n_{\mu}$=4, which represents two upward and two downward directions, which 
is a good trade-off between computational expediency and accuracy.
Both $\mathcal{J}^{\star}_{\lambda}$ and $\mathcal{J}^{\rm{d}}_{\lambda}$ vary through
the atmosphere, as seen in Fig. (\ref{plot_averdn_fig}). 
Diffuse radiation plays the fundamental role of transferring energy 
between altitudes.
\\

\subsection*{Numerical solution to the radiative transfer equation}

The solution for the non-diffuse radiance $I_{\lambda}^{\star}(\mathbf{x,s})$ is trivial
and given by Beer-Lambert's law. 
To solve for the diffuse radiance $I_{\lambda}^{\rm{d}}(\mathbf{x,s})$
 we choose to recast the radiative transfer equation as:
\begin{equation*}
\frac{dI^{\rm{d}}_{\lambda}}{d(\tau_{\lambda}/\mu_k)}=-I^{\rm{d}}_{\lambda} + S_{\lambda}, 
\label{rte2_eq}
\end{equation*}
where $S_{\lambda}=\varepsilon_{\lambda}/\kappa_{\lambda}$ is the source function, 
$\tau_{\lambda}$ is the optical thickness in the substellar direction and $\mu_{k}$ 
defines the orientation of the radiation ray (Fig. \ref{rte_kelt_sketch_fig}). 
The formal solution to this equation with a zero-influx at the boundary is:
\begin{equation*}
I^{\rm{d}}_{\lambda}=\int_{0}^{\infty} S_{\lambda}(\tau) \exp{(-\tau/\mu_{k})} d(\tau/\mu_{k}),
\end{equation*}
which represents the source function weighted by the transmittance of the gas column. 
$S_{\lambda}(\tau)$ varies slowly with $\tau$ even when the optical thickness
changes rapidly. Numerically, we approximate:
\begin{eqnarray*}
I^{\rm{d}}_{\lambda} \approx 
S_{\lambda}(\tau_1=0) \int_{\tau_1=0}^{\tau_{3/2}} \exp{(-\tau/\mu_{k})} d(\tau /\mu_{k})
+
\sum_{l>1} S_{\lambda}(\tau_l) \int_{\tau_{l-1/2}}^{\tau_{l+1/2}}   \exp{(-\tau/\mu_{k})} d(\tau/\mu_{k})
= \\
S_{\lambda}(\tau_1=0) [1-\exp(-\Delta \tau_1/2/\mu_{k})]
+
\sum_{l>1} S_{\lambda}(\tau_l) \exp{(-\tau_{l-1/2}/\mu_{k})} [1-\exp(-\Delta \tau_l/\mu_{k})]
\end{eqnarray*}
which assumes that the source function is constant within each atmospheric slab and 
where index $l$ runs over all the slabs starting from the local position along the 
specified $\mathbf{s}$ direction.
\\

The radiative transfer equation is solved over a non-uniform spectral grid of 751 bins, shown by
the green bars of Fig. (\ref{plot2_fig}). It resolves the {\lalpha} line in great detail
with a minimum bin size at the line core 
equal to 0.1$\times$ the full width at half maximum (FWHM) 
for Doppler (thermal) broadening. 
In the Lyman continuum the spectral bins are 10-50 {\angstrom} wide, whereas 
longwards of $\sim$1,500 {\angstrom} they are $\sim$300 {\angstrom} wide or larger.
We tested whether our findings were sensitive to the choice of spectral grid, and in 
particular to the details near the {\lalpha} line. 
In our tests, we always used a minimum bin size at the line core 
of (0.1--0.2)$\times$the Doppler FWHM. 
For the region $\sim$1216$\pm$300 {\angstrom}, 
we experimented with bin sizes $\Delta$$\lambda$$_{p}$ that increase towards
both the shorter and longer wavelengths, i.e.
$\Delta$$\lambda$$_{p+1}$=$\Delta$$\lambda$$_{p}$$\times$$r_{\rm{stretch}}$, where
$\Delta$$\lambda$$_{0}$ is the (minimum) bin size at the {\lalpha} core. 
We used stretching factors $r_{\rm{stretch}}$ ranging from $\sim$1.40 
(coarse grid) to $\sim$1.01 (fine grid). 
This translates into grids that contain from 100 to 750 bins. 
Over all these sensitivity experiments, the calculated mass loss rate for our fiducial 
case varied by less than 15\%. The calculations presented here use 
$r_{\rm{stretch}}$$\sim$1.01 and $\Delta$$\lambda$$_{0}$=0.1$\times$Doppler FWHM.
\\

\subsection*{Stellar irradiation}

\begin{figure*}[h]
   \centering
   \includegraphics[width=14.cm]{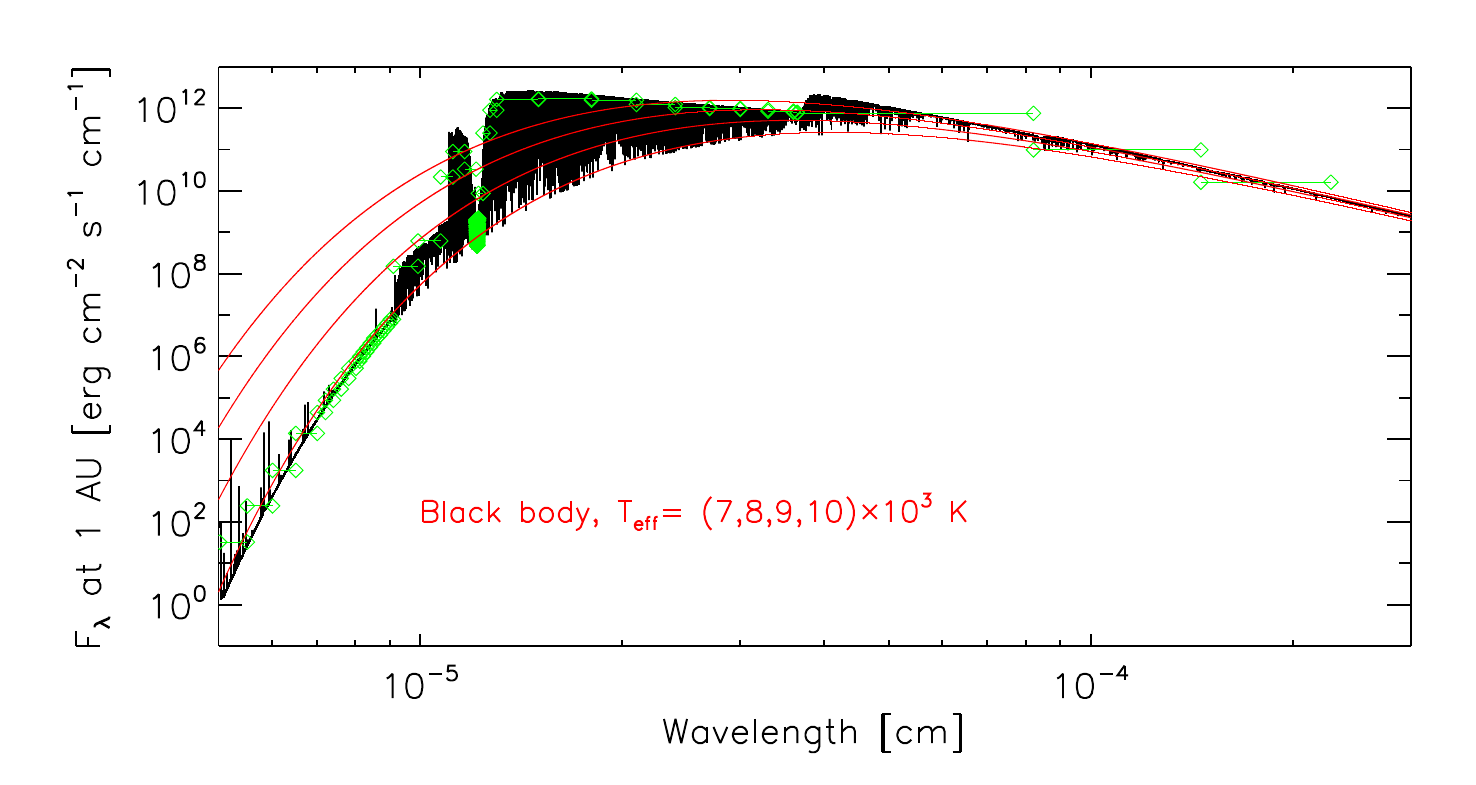}
      \caption{\label{plot2_fig} PHOENIX spectrum adopted for KELT-9 \citep{husseretal2013}. 
      For reference, black body emissions at the specified temperatures are also shown.
      The green bars show the implemented lower-resolution spectrum. 
      }
   \end{figure*}

For our model simulations, we implemented the PHOENIX spectrum shown 
in Fig. (\ref{plot2_fig}) \citep{husseretal2013}.
It has an EUV-integrated irradiance 
$F^{\star}_{\rm{LyC}}$=$\int_{\lambda<912 {\angstrom}} \mathcal{F}^{\star}_{\lambda} d\lambda$=3.8 (3,100) 
erg s$^{-1}$cm$^{-2}$ at 1 (0.035) AU. 
The EUV energy is orders of magnitude less than 
$F^{\star}_{\rm{BaC}}$=$\int_{\lambda<3646 {\angstrom}} \mathcal{F}^{\star}_{\lambda} d\lambda$=2.9$\times$10$^7$ erg s$^{-1}$cm$^{-2}$ (1 AU), 
which results in $F^{\star}_{\rm{BaC}}$/$F^{\star}_{\rm{LyC}}$$\sim$7.5$\times$10$^6$.
Figure (\ref{plot_averdn_fig}) shows the 
stellar average intensity $\mathcal{J}^{\star}_{\lambda}$ (red curves) in the adopted
spectral grid of our model.  
At very high altitudes with negligible absorption $\mathcal{J}^{\star}_{\lambda}$$\approx$
$\mathcal{F}^{\star}_{\lambda}$/4$\pi$. 
\\

Details in the stellar model such as temperature structure, chemical abundances or 
the LTE/NLTE treatment of the radiation problem will
surely affect the estimated EUV output of KELT-9. 
Our work shows though that KELT-9b's thermospheric structure is largely dictated by energy
deposition in the Balmer continuum. As a result, uncertainties in KELT-9's EUV spectrum by
a factor of up to a few have a negligible impact on the planet thermosphere.
We moreover confirm that for our fiducial model, increasing the
stellar EUV spectrum by $\times$6 produces a change in the mass loss rate 
$\dot{m}$$<$3\%.

\section*{Numerical integration}

The hydrodynamic and NLTE problems are strongly coupled and must be solved self-consistently.
In our implementation, we proceed sequentially by calculating the radiation field on the 
basis of the best estimate of atmospheric properties at the time. 
The outcome of this step is the average intensity $\mathcal{J}_{\lambda}(\mathbf{x})$
at each location along the substellar line.
Having calculated the radiation field, the hydrodynamics and NLTE equations 
are solved, enabling us to re-estimate the radiation field. 
This iterative procedure is repeated until convergence of all the variables in the
radiation, population and hydrodynamics problems.
\\

\section*{Breakdown of net energy emission rates in the fiducial model}

By construction, the net energy emission rate that goes into the energy conservation
equation is:
$$\Gamma(\mathbf{x})=\Gamma^{\rm{BB}}(\mathbf{x})+\Gamma^{\rm{BF/FB}}(\mathbf{x})+\Gamma^{\rm{FF}}(\mathbf{x}),$$
which contains contributions from BB, BF/FB and FF transitions.
\\

In our treatment of BB transitions, we assume that all lines except {\lalpha} are transparent, and thus: 
\begin{equation*}
\Gamma^{\rm{BB}}(\mathbf{x}) \approx
4\pi \int [-\kappa^{\rm{Ly-}\alpha}_{\lambda}(\mathbf{x}) \mathcal{J}_{\lambda}(\mathbf{x}) 
+ \varepsilon^{\rm{Ly-}\alpha}_{\lambda}(\mathbf{x})] d\lambda + 
\sum_{j>i, \rm{except\;Ly-}\alpha} \frac{hc}{\lambda_{ij}} 
n_j A_{ji}.
\end{equation*}

For BF/FB transitions, we consider separate contributions from each bound state with 
principal quantum number $n$:
\begin{equation*}
\Gamma^{\rm{B(n)F/FB(n)}}(\mathbf{x}) =
4\pi \int [-\kappa^{\rm{B(n)F}}_{\lambda}(\mathbf{x}) \mathcal{J}_{\lambda}(\mathbf{x}) 
+ \varepsilon^{\rm{FB(n)}}_{\lambda}(\mathbf{x})] d\lambda.
\label{lambdaBFFB_eq}
\end{equation*}
The BF/FB contributions involving $n$=1 and 2 dominate the overall energy budget.
H(1) dominates in the upper thermosphere whereas H(2) dominates in the lower thermosphere.
\\

Finally, for FF transitions:
\begin{equation*}
\Gamma^{\rm{FF}}(\mathbf{x})=4\pi \int [-\kappa^{\rm{FF}}_{\lambda}(\mathbf{x}) \mathcal{J}_{\lambda}(\mathbf{x}) 
+ \varepsilon^{\rm{FF}}_{\lambda}(\mathbf{x})] d\lambda.
\end{equation*}


\clearpage


\begin{thebibliography}{99}

\bibitem[Ballester \& Ben-Jaffel(2015)]{ballesterbenjaffel2015}
Ballester, G.E. \& Ben-Jaffel, L.\ 2015,
\apj, 804:116

\bibitem[Ben-Jaffel(2007)]{benjaffel2007}
Ben-Jaffel, L.\ 2007,
\apj, 671:L61-L64

\bibitem[Bourrier et al.(2018)]{bourrieretal2018}
Bourrier, V., 
Lecavelier des Etangs, A., 
Ehrenreich, D., 
Sanz-Forcada, J., 
Allart, R., et al.\ 2018,
A\&A, 620:A147

\bibitem[Cauley et al.(2019)]{cauleyetal2019}
Cauley, P.W., 
Shkolnik, E.L., 
Ilyin, I., 
Strassmeier, K.G., 
Redfield, S. \& Jensen, A.\ 2019,
\aj, 157:69

\bibitem[Casasayas-Barris et al.(2018)]{casasayasbarrisetal2018}
Casasayas-Barris, N., 
Pall\'e, E., 
Yan, F., 
Chen, G., 
Albrecht, S., et al.\ 2018,
A\&A, 616:A151

\bibitem[Christie et al.(2013)]{christieetal2013}
Christie, D.,
Arras, P. \& Li, Z.-Y.\ 2013,
\apj, 772:144

\bibitem[Collier Cameron et al.(2010)]{colliercameronetal2010}
Collier Cameron, A., 
Guenther, E., 
Smalley, B., 
McDonald, I., 
Hebb, L., et al.\ 2010,
\mnras, 407:507

\bibitem[Debrecht et al.(2019)]{debrechtetal2019}
Debrecht, A., 
Carroll-Nellenback, J., 
Frank, A., 
McCann, J., 
Murray-Clay, R. \& Blackman, E.G.\ 2019,
\mnras, 483:1481--1495

\bibitem[Ehrenreich et al.(2015)]{ehrenreichetal2015}
Ehrenreich, D.,
Bourrier, V., 
Wheatley, P.J., 
Lecavelier des Etangs, A., 
H\'ebrard, G.\ 2015,
Nature, 522:459

\bibitem[Fossati et al.(2010)]{fossatietal2010}
Fossati, L., 
Haswell, C.A., 
Froning, C.S., 
Hebb, L., 
Holmes, S., et al.\ 2010,
\apjl, 714:L222

\bibitem[Fossati et al.(2018)]{fossatietal2018}
Fossati, L.,
Koskinen, T.,
Lothringer, J.D., 
France, K.,
Young, M.E. \& 
Sreejith, A.G.\ 2018,
\apjl, 868:L30

\bibitem[Fulton et al.(2017)]{fultonetal2017}
Fulton, B.J., 
Petigura, E.A., 
Howard, A.W., 
Isaacson, H., 
Marcy, G.W., et al.\ 2017,
\apj, 154:109

\bibitem[Garc\'ia Mu\~noz(2007)]{garciamunoz2007}
Garc\'ia Mu\~noz, A.\ 2007,
Planet. Space Science, 55:1426

\bibitem[Gaudi et al.(2017)]{gaudietal2017}
Gaudi, B.S., 
Stassun, K.G., 
Collins, K.A.,
Beatty, T.G., 
Zhou, G., et al.\ 2017,
Nature, 546:514

\bibitem[Gnat \& Ferland(2002)]{gnatferland2012}
Gnat, O. \& Ferland, G.J.\ 2012,
\apjs, 199:20

\bibitem[Guo \& Ben-Jaffel(2016)]{guobenjaffel2016}
Guo, J.H. \& Ben-Jaffel, L.\ 2016,
\apj, 818:107

\bibitem[Hoeijmakers et al.(2018)]{hoeijmakersetal2018}
Hoeijmakers, H.J., 
Ehrenreich, D., 
Heng, K., 
Kitzmann, D.,
Grimm, S.L., et al.\ 2018,
Nature, 560:453

\bibitem[Hoeijmakers et al.(2019)]{hoeijmakersetal2019}
Hoeijmakers, H.J., 
Ehrenreich, D., 
Kitzmann, D., 
Allart, R., 
Grimm, S.L., et al.\ 2019,
A\&A, \textit{in press}

\bibitem[Huang et al.(2017)]{huangetal2017}
Huang, C.,
Arras, P.,
Christie, D. \&
Li, Z.-Y.\ 2017,
\apj, 851:150

\bibitem[Husser et al. (2013)]{husseretal2013}
Husser, T.-O., 
Wende-von Berg, S., 
Dreizler, S., 
Homeier, D., 
Reiners, A., et al.\ 2013,
A\&A, 553:A6

\bibitem[Ionov et al.(2014)]{ionovetal2014}
Ionov, D.E., 
Bisikalo, D.V., 
Shematovich, V.I. \& Huber, B.\ 2014,
Solar System Research, 48:105

\bibitem[Jensen et al.(2012)]{jensenetal2012}
Jensen, A.G., 
Redfield, S., 
Endl, M., 
Cochran, W.D., 
Koesterke, L. \& Barman, T.\ 2012,
\apj, 751:86

\bibitem[Jin \& Mordasini(2018)]{jinmordasini2018}
Jin, S. \& Mordasini, C.\ 2018,
\apj, 853:163

\bibitem[Kitzmann et al.(2018)]{kitzmannetal2018}
Kitzmann, D.,
Heng, K.,
Rimmer, P.B.,
Hoeijmakers, H.J.,
Tsai, S.-M., et al.\ 2018,
\apj, 863:183



\bibitem[Koskinen et al.(2013)]{koskinenetal2013}
Koskinen, T.T., 
Harris, M.J., 
Yelle, R.V. \& Lavvas, P.\ 2013,
Icarus, 226, 2, 1678--1694

\bibitem[Kulow et al.(2014)]{kulowetal2014}
Kulow, J.R., 
France, K., 
Linsky, J. \& Loyd, R.O.P.\ 2014,
\apj, 786:132

\bibitem[Lammer et al.(2003)]{lammeretal2003}
Lammer, H., 
Selsis, F., 
Ribas, I., 
Guinan, E.F., 
Bauer, S.J. \& Weiss, W.W.\ 2003,
\apj, 598:L121

\bibitem[Lammer et al.(2008)]{lammeretal2008}
Lammer, H., 
Kasting, J.F., 
Chassefi\`ere, E., 
Johnson, R.E., 
Kulikov, Y.N. \& Tian, F.\ 2008,
\ssr, 139:399

\bibitem[Lopez et al.(2012)]{lopezetal2012}
Lopez, E.D., 
Fortney, J.J. 
\& Miller, N.\ 2012,
\apj, 761:59

\bibitem[Lothringer et al.(2018)]{lothringeretal2018}
Lothringer, J.D.,
Barman, T. \& Koskinen, T.\ 2018,
\apj, 866:27

\bibitem[Lecavelier des Etangs et al.(2010)]{lecavelierdesetangsetal2010}
Lecavelier Des Etangs, A., 
Ehrenreich, D., 
Vidal-Madjar, A., 
Ballester, G.E., 
D\'esert, J.-M., et al.\ 2010, 
A\&A, 514:A72

\bibitem[Linsky et al.(2010)]{linskyetal2010}
Linsky, J.L., 
Yang, H., 
France, K., 
Froning, C.S., 
Green, J.C., et al.\ 2010,
\apj, 717:1291

\bibitem[Lund et al.(2017)]{lundetal2017}
Lund, M.B., 
Rodriguez, J.E., 
Zhou, G., 
Gaudi, B.S., 
Stassun, K.G., et al.\ 2017,
\aj, 154:194

\bibitem[Menager et al.(2013)]{menageretal2013}
Menager, H.,
Barth\'elemy, M.,
Koskinen, T.,
Lilensten, J.,
Ehrenreich, D. \& 
Parkinson, C.D.\ 2013,
Icarus, 226:1709

\bibitem[Munaf\`o et al.(2017)]{munafoetal2017}
Munaf\`o, A.,
Mansour, N.N. \&
Panesi, M.\ 2017,
\apj, 838:126

\bibitem[Murray-Clay et al.(2009)]{murrayclayetal2009}
Murray-Clay, R.A., 
Chiang, E.I. \& Murray, N.\ 2009,
\apj, 693:23

\bibitem[Owen(2019)]{owen2019}
Owen, J.E.\ 2019,
Annu. Rev. Earth Planet. Sci., 47:67

\bibitem[Salz et al.(2016)]{salzetal2016}
Salz, M., 
Czesla, S., 
Schneider, P.C. \& Schmitt, J.H.M.M.\ 2016, 
A\&A, 586:A75

\bibitem[Salz et al.(2018)]{salzetal2018}
Salz, M., 
Czesla, S., 
Schneider, P.C., 
Nagel, E., 
Schmitt, J.H.M.M., et al.\ 2018,
A\&A, 620:A97

\bibitem[Sch\"oll et al.(2016)]{schoelletal2016}
Sch\"oll, M., 
Dudok de Wit, T., 
Kretzschmar, M. \&
Haberreiter, M.\ 2016,
J. Space Weather Space Clim., 6:A14

\bibitem[Shaikhislamov et al.(2018)]{shaikhislamovetal2018}
Shaikhislamov, I.F., 
Khodachenko, M.L., 
Lammer, H., 
Berezutsky, A.G., 
Miroshnichenko, I.B. \& Rumenskikh, M.S.\ 2018,
\mnras, 481:5315

\bibitem[Sing et al.(2019)]{singetal2019}
Sing, D.K., 
Lavvas, P.,
Ballester, G.E., 
Lecavelier des Etangs, A., 
Marley, M.S., et al.\ 2019,
\aj, 158:91

\bibitem[Spake et al.(2018)]{spakeetal2018}
Spake, J.J., 
Sing, D.K., 
Evans, T.M., 
Oklop{\v c}i\'c, A.,
Bourrier, V., et al.\ 2018,
Nature, 557:68

\bibitem[Talens et al.(2018)]{talensetal2018}
Talens, G.J.J., 
Justesen, A.B., 
Albrecht, S., 
McCormac, J., 
Van Eylen, V., et al.\ 2018,
A\&A, 612:A57

\bibitem[Thuillier et al.(2004)]{thuillieretal2004}
Thuillier, G., 
Floyd, L., 
Woods, T.N.,  
Cebula, R.,
Hilsenrath, E., 
Hers\'e, M., \&
Labs, D.\ 2004,   
Solar irradiance reference spectra. 
In: J.M. Pap, P. Fox, C. Frohlich, H.S. Hudson, J. Kuhn, J. McCormack, 
G. North, W. Sprigg, and S.T. Wu, Editors. 
Solar variability and its effects on climate, 
Geophysical Monograph 141, American Geophysical Union, 
Washington, DC, 171.

\bibitem[Tian et al.(2005)]{tianetal2005}
Tian, F., 
Toon, O.B., 
Pavlov, A.A. \& De Sterck, H.\ 2005,
\apj, 621:1049

\bibitem[Tian(2015)]{tian2015}
Tian, F.\ 2015,
Annu. Rev. Earth Planet. Sci., 43:459

\bibitem[Trammell et al.(2014)]{trammelletal2014}
Trammell, G.B., 
Li, Z.-Y. \& Arras, P.\ 2014,
\apj, 788:161

\bibitem[Tremblin \& Chiang(2013)]{tremblinchiang2013}
Tremblin, P. \& Chiang, E.\ 2013,
\mnras, 428:2565

\bibitem[Vidal-Madjar et al.(2003)]{vidalmadjaretal2003}
Vidal-Madjar, A., 
Lecavelier des Etangs, A., 
D\'esert, J.-M., 
Ballester, G.E., 
Ferlet, R., H\'ebrard, G. \& Mayor, M.\ 2003,
Nature, 422:143

\bibitem[Vidal-Madjar et al.(2004)]{vidalmadjaretal2004}
Vidal-Madjar, A., 
D\'esert, J.-M., 
Lecavelier des Etangs, A., 
H\'ebrard, G., 
Ballester, G.E., et al.\ 2004,
\apj, 604:L69

\bibitem[Yan \& Henning(2018)]{yanhenning2018}
Yan, F. \& Henning, T.\ 2018,
Nature Astronomy, 2:714

\bibitem[Wang et al.(2014)]{wangetal2014}
Wang, Y.,
Ferland, G.J.,
Lykins, M.,
Porter, R.,
van Hoof, P.A.M. \& Williams, R.J.R.\ 2014,
\mnras, 440:3100

\bibitem[Yelle(2004)]{yelle2004}
Yelle, R.V.\ 2004,
Icarus, 170:167

\bibitem[Zahnle \& Catling(2017)]{zahnlecatling2017}
Zahnle, K.J. \& Catling, D.C.\ 2017, 
\apj, 843:122

\end{thebibliography}

\newpage

\begin{thebibliography}{99}

\bibitem[Le Teuff et al.(2000)]{leteuffetal2000}
Le Teuff, Y.H.,
Millar, T.J. \& 
Markwick, A.J.\ 2000,
A\&A Suppl. Ser., 146, 157

\bibitem[Bates et al.(1962a)]{batesetal1962a}
Bates, D. R., 
Kingston, A.E. \& McWhirter, R.W.P\ 1962,
RSPSA, 267, 297

\bibitem[Bates et al.(1962b)]{batesetal1962b}
Bates, D. R., 
Kingston, A.E. \& McWhirter, R.W.P\ 1962,
RSPSA, 270, 155

\bibitem[Kramida(2010)]{kramida2010}
Kramida, A.\ 2010,
Atomic Energy Levels and Spectra Bibliographic Database (version 2.0). 
Available: https://physics.nist.gov. 
DOI: 10.18434/T40K53.

\bibitem[Anderson et al.(2000)]{andersonetal2000}
Anderson, H.,
Ballance, C.P.,
Badnell, N.R. \& 
Summers, H.P.\ 2000,
J. Phys. B: At. Mol. Opt. Phys., 33, 1255

\bibitem[Anderson et al.(2002)]{andersonetal2002}
Anderson, H.,
Ballance, C.P.,
Badnell, N.R. \& 
Summers, H.P.\ 2002,
J. Phys. B: At. Mol. Opt. Phys., 35, 1613

\bibitem[Przybilla \& Butler(2004)]{przybillabutler2004}
Przybilla, N. \&
Butler, K.\ 2004,
ApJ, 09:1181

\bibitem[Aggarwal et al.(2018)]{aggarwaletal2018}
Aggarwal, K.M.,
Owada, R. \& 
Igarashi, A.\ 2018,
Atoms, 6, 37

\bibitem[Barklem(2007)]{barklem2007}
Barklem, P.S.\ 2007,
A\&A, 466, 327

\bibitem[Vriens \& Smeets(1980)]{vrienssmeets1980}
Vriens, L. \& Smeets, A.H.M.\ 1980,
Phys. Rev. A, 22, 940

\bibitem[Seaton(1955)]{seaton1955}
Seaton, M.J.\ 1955,
Proc. Phys. Soc. A, 68, 457

\bibitem[Struensee \& Cohen(1988)]{struenseecohen1988}
Struensee, M.C. \& Cohen, J.S.\ 1988,
Phys. Rev. A, 38, 3377

\bibitem[Colonna et al.(2012)]{colonnaetal2012}
Colonna, G.,
Pietanza, L.D. \&
D'Ammando, G.D.\ 2012,
Chem. Phys., 398, 37

\bibitem[Whiting(1968)]{whiting1968}
Whiting, E.E.\ 1968,
JQSRT, 8, 1379

\bibitem[Griem(1974)]{griem1974}
Griem, H.R.\ 1974,
Spectral Line Broadening by Plasmas, Academic Press, New York and London. 

\bibitem[Kramida et al.(2018)]{kramidaetal2018}
Kramida, A., 
Ralchenko, Yu., 
Reader, J., \& NIST ASD Team\ 2018. 
NIST Atomic
Spectra Database (ver. 5.5.6), [Online]. Available: https://physics.nist.gov/asd. 
DOI: https://doi.org/10.18434/T4W30F

\bibitem[Nussbaumer \& Schmutz(1984)]{nussbaumerschmutz1984}
Nussbaumer, H. \& Schmutz, W.\ 1984,
A\&A, 138, 495

\bibitem[Dennison et al.(2005)]{dennisonetal2005}
Dennison, B., Turner, B.E. \& Minter, A.H.\ 2005,
ApJ, 633, 309

\bibitem[Mihalas(1978)]{mihalas1978}
Mihalas, D.\ 1978,
Stellar Atmospheres, 2nd Edition, W.H. Freeman and Company, San Francisco.

\end{thebibliography}
\end{document}